\documentclass[nofootinbib,a4paper,aps,prd,10pt,superscriptaddress,reprint,showkeys,showpacs,twocolumn]{revtex4}

\usepackage{graphicx}
\usepackage{amsmath}
\usepackage{amsfonts}
\usepackage{amsthm}
\usepackage{mathrsfs}
\usepackage{amssymb}
\usepackage{epsfig}
\usepackage{amscd}
\usepackage{dcolumn}
\usepackage{bm}
\usepackage{natbib}
\usepackage{url}
\usepackage{xspace}

\usepackage[normalem]{ulem}

\usepackage{xcolor}

\usepackage{hyperref}
\hypersetup{%
    ,urlcolor=blue
    ,citecolor=blue
    ,linkcolor=blue
    }

\def\be{\begin{equation}}
\def\ee{\end{equation}}
\def\bea{\begin{eqnarray}}
\def\eea{\end{eqnarray}}

\def\nat{Nature}
\def\prl{Phys. Rev. Lett.}

\def\prd{Phys. Rev. D}

\def\mnras{Mon. Not. Roy. Astr. Soc.}
\def\apj{Astrophys. J.}
\def\apjl{Astrophys. J. Lett.}

\def\aap{Astron. Astrophys.}

\def\pasj{Publications of the Astronomical Society of Japan }

\def\physrep{Phys. Rep.}
\def\jcap{Journal of Cosmology and Astroparticle Physics}

\begin{document}

\title{Effects of non-vanishing dark matter pressure in the Milky Way galaxy}

\author{Kuantay Boshkayev}
\email{kuantay@mail.ru}
\affiliation{National Nanotechnology Laboratory of Open Type,  Almaty 050040, Kazakhstan.}
\affiliation{Al-Farabi Kazakh National University, Almaty 050040, Kazakhstan.}
\affiliation{Department of Engineering Physics, Satbayev University, 22 Satbayev str., 050013 Almaty, Kazakhstan.}

\author{Talgar Konysbayev}
\email{talgar\_777@mail.ru}
\affiliation{National Nanotechnology Laboratory of Open Type,  Almaty 050040, Kazakhstan.}
\affiliation{Al-Farabi Kazakh National University, Almaty 050040, Kazakhstan.}

\author{Ergali Kurmanov}
\email{kurmanov.yergali@kaznu.kz}
\affiliation{National Nanotechnology Laboratory of Open Type,  Almaty 050040, Kazakhstan.}
\affiliation{Al-Farabi Kazakh National University, Almaty 050040, Kazakhstan.}

\author{Orlando Luongo}
\email{orlando.luongo@unicam.it}
\affiliation{Scuola di Scienze e Tecnologie, Universit\`a di Camerino, Via Madonna delle Carceri 9, 62032 Camerino, Italy.}
\affiliation{Dipartimento di Matematica, Universit\`a di Pisa, Largo B. Pontecorvo 5, Pisa, 56127, Italy.}
\affiliation{NNLOT, Al-Farabi Kazakh National University, Al-Farabi av. 71, 050040 Almaty, Kazakhstan.}

\author{Daniele Malafarina}
\email{daniele.malafarina@nu.edu.kz}
\affiliation{Department of Physics, Nazarbayev University, Astana 010000, Kazakhstan.}

\author{Kalbinur Mutalipova}
\email{kalbinur.mutalipova@mail.ru}
\affiliation{Al-Farabi Kazakh National University, Almaty 050040, Kazakhstan.}

\author{Gulnur Zhumakhanova}
\email{zhumakhanovag@gmail.com}
\affiliation{Al-Farabi Kazakh National University, Almaty 050040, Kazakhstan.}

\begin{abstract}
We consider the possibility that the Milky Way's dark matter halo possesses a non vanishing equation of state. Consequently, we evaluate the contribution due to the speed of sound, assuming that the dark matter content of the galaxy behaves like a fluid with pressure. In particular, we model the dark matter distribution via an exponential sphere profile in the galactic core, and inner parts of the galaxy whereas we compare the exponential sphere with three widely-used profiles for the halo, i.e. the Einasto, Burkert and Isothermal profile. For the galactic core we also compare the effects due to a dark matter distribution without black hole with the case of a supermassive black hole in vacuum and show that present observations are unable to distinguish them. Finally we investigate the expected experimental signature provided by gravitational lensing due to the presence of dark matter in the core.
\end{abstract}



\maketitle

\section{Introduction}\label{introduzione}

The structure, morphology and evolution of galaxies are currently based on two important assumptions that are part of the standard model of galaxy formation. The first requirement is the  presence of super-massive black hole (SMBH) at the center of almost every spiral galaxy  \cite{formazionegalassie}. The second is the unavoidable existence of dark matter (DM) halos  surrounding every galaxy \cite{Nakama2020}. In particular, DM existence was originally proposed to explain structure formation and to explain the observed rotational curves (RCs) of stars at the periphery of galaxies. The existence of DM is today strongly confirmed by several  independent observations  \cite{Freese2017}. Although such observations suggest that DM does exist, its nature, micro-physics and fundamental properties are still unknown, and so far no suitable DM candidates have been directly detected \cite{orl1,orl2,orl3,orl4}.

The observed RCs of galaxies other than the Milky Way are modelled by making use of phenomenological DM profiles \cite{Cirelli2011}, whereas for the Milky Way galaxy (MWG), the available data on the RCs of stars at different distances from the center lead to a more complicated paradigm with different DM profiles depending on the distance from the center \cite{Siutsou2015,Nesti2013}.

On the other hand, the hypothesis of core-placed SMBHs is needed to explain the powerful X-ray radiation observed from quasars and the nature of Active Galactic Nuclei (AGN). Observations of SMBHs up to distances of redshift $z=7.54$ show that they had already formed less than one billion years after the Big Bang \cite{Ba2018}. Unfortunately, it is currently unclear how such enormously massive black holes could have developed so quickly. Several, less active, SMBHs have also been detected at closer distances. Most notably, a SMBH candidate has been observed via direct imaging at the core of the galaxy M87 \cite{EHT} and regarding our own MWG, observations of stellar dynamics near the galactic center suggest the existence of a massive compact object located in the constellation of Sagittarius, named Sagittarius-A$^\star$, or briefly Sgr-A$^\star$, with mass $M_{Sgr^\star}\simeq 4\cdot 10^6\,M_\odot$ (with $M_\odot$ being the Sun's mass)\cite{2000ApJ...528L..13F}. Constraints on its size made via Very-Long-Baseline-Interferometry in the millimeter wavelengths give an estimate of less than $0.5$ astronomical unit (AU) \cite{2018ApJ...859...60L}, with the closest observed star, called S2, orbiting Sgr-A$^\star$ with a distance at peribothron of roughly $120$ AU.

We know that the gravitational field of the SMBHs must dominate from the center of the galaxy to a distance of a few parsecs, while at larger distances DM and stellar densities prevail \cite{Sofue2013}. It is reasonable to suppose that DM distributions must extend to the center of galaxies and even though the role of DM surrounding SMBHs is currently under debate \cite{2020MNRAS.496.1115B,galaxies2020} it seems reasonable to assume that DM envelopes must surround SMBHs.
In addition, from a cosmological perspective, it appears possible that DM may not be in the form of dust \cite{mio1,mio2,mio3}, i.e. pressureless matter. The assumption of non vanishing pressures for DM then may bear important consequences both for cosmology  as well as for the structure of galaxies \cite{orla1,orla2,orla3,orla4,mio4,mio5}.

In this paper, we assume that DM is not pressureless and it gravitationally forms bound clumps over fairly short time scales and investigate qualitatively how a model of DM with pressure could account for observations.
In particular, we focus our attention to the MWG, where we have a wide number of observations and consider a viable DM distribution from the halo to the center of the galaxy. We characterize the different contributions of DM at different distances from the center through the exponential sphere density profile. This profile is well suited for the core, since it does not diverge at the center. Also, for the halo of the galaxy, we consider several widely used density profiles for DM and show that the exponential sphere is well suited to reproduce the data.

Our main goals are to (i) compute theoretically the properties that a DM distribution would exhibit under the hypothesis of a non-vanishing equation of state and (ii) compare such properties with observations in the MWG. In particular we are interested in understanding the role played by pressure in the motion of stars in the MWG and the possibility that the central compact object be a SMBH of smaller mass surrounded by a massive DM envelope.
To this aim we proceed with the following steps:
\begin{itemize}
\item[{\bf I.}] We first analyse RCs of the MWG to infer the values of physical parameters necessary for the description of the DM profiles.
\item[{\bf II.}] We derive the equations of state for DM that correctly reproduce the observations at various distances from the galactic center.
\item[{\bf III.}] From the equation of state we obtain the sound speed in DM for each profile and discuss its implications.
\item[{\bf IV.}] Considering the core of the galaxy, we compare the lensing effect in the two extreme cases of only DM without SMBH and only SMBH without DM.
\end{itemize}

The aim of the last point is to suggest that current observations of M87 and Sgr-A$^\star$ can not rule out that the observed affects (i.e. shadow and motion of stars) may be due to a combination of a DM profile surrounding a smaller SMBH.
We expect that future observations of motion of stars near the galactic center and light emitted by the gas in the vicinity of the SMBH in M87 will allow us to better constrain the mass ratio between SMBH and DM. In particular, when applied to distant AGNs and quasars, the idea implies that such sources may be in fact be less massive SMBHs immersed in denser DM envelopes.

The paper is organized as follows. In Sect.~\ref{goal}, the model is described, RCs are discussed and the equations for the DM profiles at various distances from the center are presented.
In Sect.~\ref{eos}, we obtain constraints on the DM equation of state and pressure in the MWG and discuss the speed of sound for DM in the proposed model.

In Sect.~\ref{refindex} we focus on the galactic core and compare the effects due to lensing in the case of a SMBH in vacuum with those due to a DM distribution without the black hole. We show that significant differences in the lensing appear at short distances and that future observations may be able to provide constraints to the DM to SMBH mass ratio.
Finally, Sect.~\ref{concl} is devoted to a brief discussion of possible future observations that may allow to test the ideas proposed here.


\section{Modelling the Milky Way's rotation curves}
\label{goal}

To model RCs in spirals we may use available data on star's velocities. Unfortunately for galaxies other than the MWG the only available data is for stars in the outer regions of the galaxy. On the other hand, for the MWG we have data for stars at various distances from the galactic center, which allows us to reconstruct the DM profile for the whole galaxy.

Therefore the following model is generic enough to be applied to any galaxy but the specific DM density profiles are obtained from data on star's velocities in the MWG.

For the galaxy's matter content we consider a DM profile that extends throughout the galaxy and is divided into several components, i.e. core, bulge, disk and halo.
We then tailor the parameters of the models to reproduce the behavior observed in the MWG.
From observational data of galactic RCs, we then calculate the mass of each part of the galaxy in our model.
In this respect, this is performed by using phenomenological density profiles for DM under the hypothesis that DM has non zero pressure, i.e. assuming a non vanishing equation of state that is obtained from hydrostatic equilibrium equations.

Additionally we may consider also a SMBH located at the galactic center. As expected, the contribution of the SMBH is dominant only very close to the galactic center and therefore will become relevant only when dealing with stars in the core.
Thus, for the core, in analogy to what proposed in Ref.~\cite{Boshkayev2019}, we concentrate on fitting the existing observational data for the MWG with the following two `extreme' theoretical cases:
\begin{itemize}
\item[{\bf 1.}] There is a SMBH at the center of the galaxy and the DM distribution in the core is negligible.
\item[{\bf 2.}] There is a clump of DM at the center of the galaxy, whose total mass is approximately equal to that of the SMBH candidate, and there is no SMBH.
\end{itemize}
By exploring the observational consequences of these two extreme possibilities we argue that any situation sitting in between (i.e. a central mass distribution due partly to a SMBH and partly to a DM envelope) would be indistinguishable from current available observations.

As said, different density profiles are considered to model the DM distribution in different parts of the galaxy. In particular profiles with a cusp at the center (such as the widely used Navarro-Frenk-White (NFW) \cite{Sofue2013}) will
not be considered in view of the density's diverging at the center.


\subsection{Galactic rotation curves}

Assuming a galaxy to approximately be in thermodynamic equilibrium and neglecting the contribution due to normal matter, one can proceed from general relativity (GR) to employ a spherically symmetric space-time as a first approximation to describe its DM distribution.

Thus, we can consider the following line element:
\be
\ ds^2=e^{2\Phi(r)/c^2}c^2dt^2-\frac{dr^2}{{1}-\frac{2GM(r)}{c^2r}}-{r^2d\Omega^2},\
\label{eq:sample1}
\ee
where $(t, r, \theta, \phi)$ are time and spherical coordinates and $
\ d{\Omega^2}=d\theta^2+\sin^2\theta d\phi^2$ is the line element on the unit 2-sphere. The metric functions $\Phi(r)$ and $M(r)$ are determined from Einstein's equations. Of course relativistic effects become relevant only at high densities, therefore it is useful to compare the relativistic picture with Newtonian gravity (NG) to see at what scales appreciable differences appear.

Einstein's equations in the presence of matter for the metric in Eq.~\eqref{eq:sample1} reduce to the hydrostatic equilibrium equations, known in the literature as the Tolman-Oppenheimer-Volkoff (TOV) equations, which can be written as
\bea
\frac{d M(r)}{dr}&=&4\pi r^2 \rho(r),\label{eq:sample3}\\
\frac{d P(r)}{dr}&=&-\frac{G\left(\rho(r)+\frac{P(r)}{c^2}\right)\left(M(r)+\frac{4\pi
r^3P(r)}{c^2}\right)}{r\left(r-\frac{2GM(r)}{c^2}\right)},\label{eq:sample4}\\
\frac{d\Phi(r)}{dr}&=&\frac{G\left(M(r)+\frac{4\pi
r^3P(r)}{c^2}\right)}{r\left(r-\frac{2GM(r)}{c^2}\right)}, \label{eq:sample5}
\eea
where $\rho$ and $P$ are the matter's density and pressure appearing in the energy-momentum tensor of an isotropic perfect fluid, $M$ describes the mass profile, or the amount of matter enclosed within a sphere of radius $r$ and $\Phi$ is the gravitational potential. Notice that $P$ is in general \emph{non-zero} and can be used to provide an equation of state for the matter distribution.

The assumption of a non-zero pressure, i.e. a non zero equation of state, for DM is something very different from the standard paradigm that considers DM having the form of dust. However, even very small pressures may have considerable effects. Of course, a non-vanishing DM pressure, as present in the TOV equations, is possible also for Newtonian gravity (NG).

In both cases the interpretation of DM with pressure is simple: globally the pressure is negligible, but locally its contribution cannot be exactly zero as a consequence of interacting DM constituents. Namely, weakly interacting particles can provide a small but non-zero pressure.

Assuming DM to be made of non-relativistic particles, at large distances from the center the effects of GR become negligible and we may use Newtonian physics to describe the profiles. In the limit of weak fields and small velocities, the TOV equations are reduced to the equations of hydrostatic equilibrium in NG, i.e.
\bea
\frac{d M(r)}{dr}&=&4\pi r^2 \rho(r),\label{eq:sample6}\\
\frac{d P(r)}{dr}&=&-\rho(r)\frac{GM(r)}{r^2},\label{eq:sample7}\\
\frac{d\Phi(r)}{dr}&=&\frac{GM(r)}{r^2}. \label{eq:sample8}
\eea
Then from the following relation, valid for circular orbits,
\be
\frac{d\Phi(r)}{dr}=\frac{v^2}{r}, \label{eq:sample9}
\ee
one gets the linear velocity of a test particle in the gravitational field of the mass distribution $M(r)$ in GR as
\be \label{eq:sample10}
v_{DM}=
\sqrt{\frac{\dfrac{G M(r)}{r}\left(1+\dfrac{4\pi r^3 P(r)}{c^2 M(r)}\right)}{1-\dfrac{2 G M(r)}{c^2 r}}},\,
\ee
which for NG reduces to
\be \label{eq:sample10b}
v_{DM}=
\sqrt{\dfrac{GM(r)}{r}}.\,
\ee

Similarly, one can write the linear velocity of a test particle in the field of a point particle (i.e. a black hole) of mass $M_{BH}$ in GR as
\be\label{eq:sample11}
v_{BH}=
\sqrt{\frac{\dfrac{G M_{BH}}{r}}{1-\dfrac{2 G M_{BH}}{c^2 r}}},\,
\ee
which in NG reduces to the usual equation for the orbital velocity around a point mass $M_{BH}$, given by
\be\label{eq:sample11b}
v_{BH}=
\sqrt{\dfrac{GM_{BH}}{r}}.\,
\ee
Of course, for large $r$, $c\rightarrow+\infty$ and $P/c^2\ll\rho$ the relativistic equations reduce to their Newtonian counterparts.

The linear velocities obtained from the theoretical models with given matter distributions for the galaxy can then be compared with measurements of the velocity of stars to constrain the parameters of the DM density profile in a given region of the galaxy. In the case of the MWG such measurements are available for stars located in different parts. Ranging from stars in the outer edges of the disk to stars as close as about $120$ AU from the galactic center.
Notice that in order to be able to distinguish the presence of a SMBH from a DM cloud at the galactic center we would need data on linear velocities of objects closer to the center, which is presently not available.
In the following, given the wide range of scales considered, the radial coordinate $r$ will be expressed in various units of measurement, such as AU, pc, kpc, as necessary.

\begin{figure*}[ht]
\centering
\begin{tabular}{lr}
\includegraphics[width=0.49\linewidth]{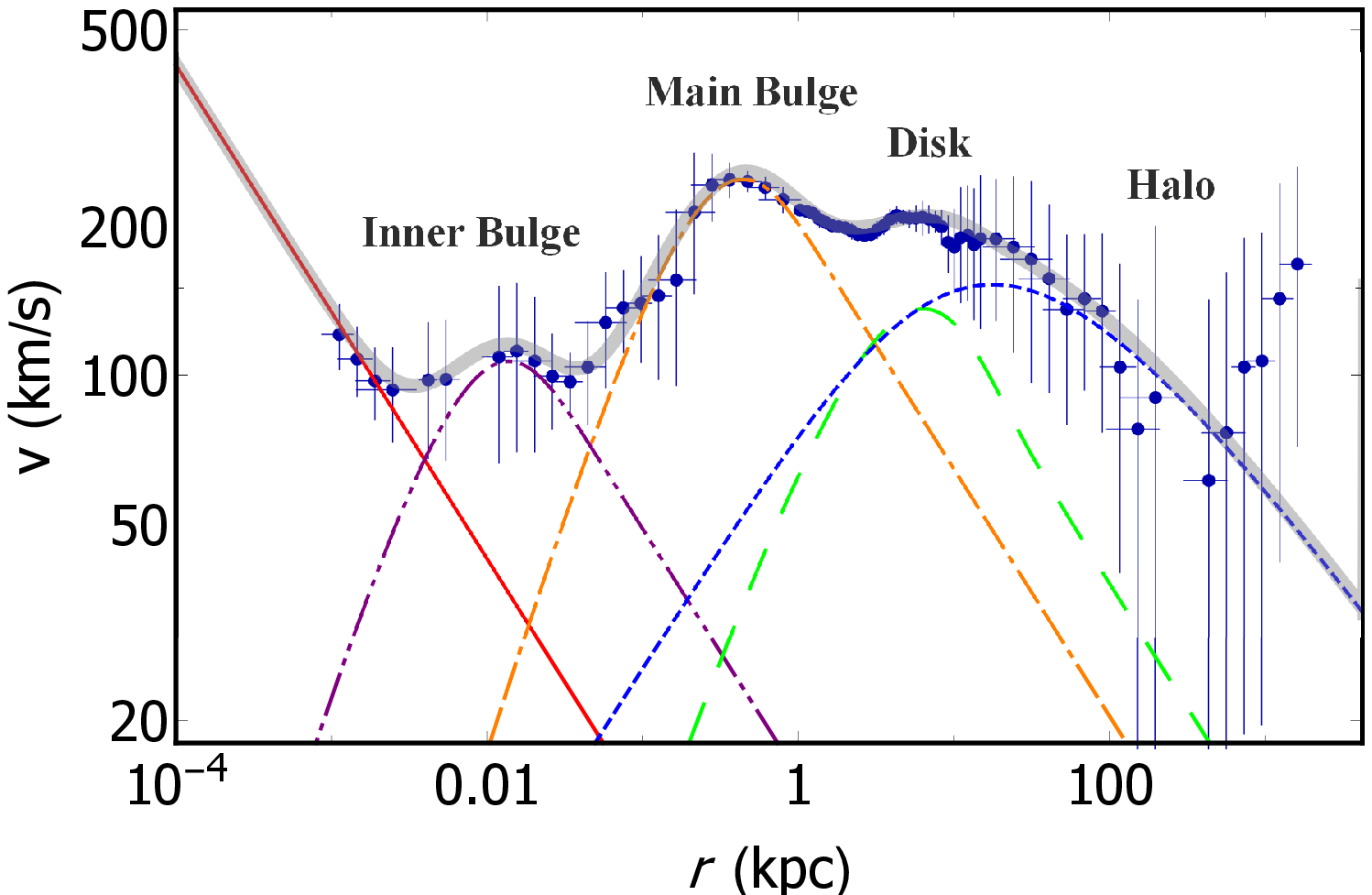}\label{ris:image1}&
\includegraphics[width=0.49\linewidth]{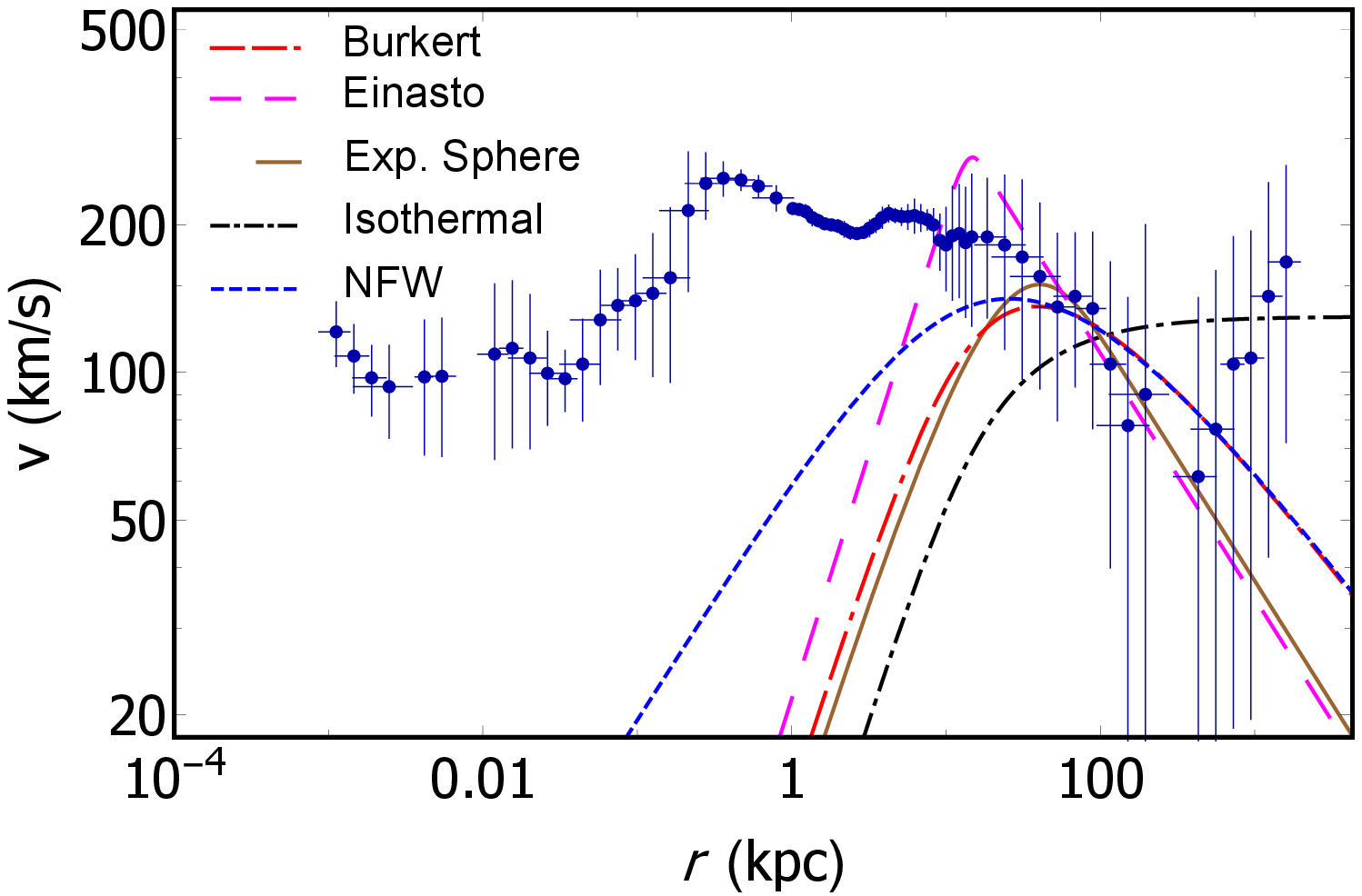}\label{ris:image1b}
\end{tabular}
\caption{Color online. Left panel: RCs of the MWG reproduced from Ref.~\cite{Sofue2013} for the entire galaxy. The central SMBH with mass $M_{BH}=4.2 \times 10^{6} M_{\odot}$ is shown by a solid red curve, the inner bulge with the exponential sphere density proﬁle is shown by a dashed and double dotted purple curve, the main bulge with the exponential sphere density proﬁle is shown by a double dashed and dotted orange curve, the disk is shown by a dashed green curve and the halo described by the Navarro-Frenk-White (NFW) proﬁle \cite{Castignani2012,Navarro} is shown by a dashed blue curve.
Right panel: RC of the halo of the MWG with different density profiles. The various curves represent the best fits for the DM halo using four different profiles, namely Burkert, Einasto, exponential sphere, isothermal and NFW. Notice that the isothermal profile is significantly different from the others and the behavior of the mass function at large distances suggests that it is less suitable to model the MWG's halo.}
\label{ris:image1ab}
\end{figure*}

\begin{table*}[ht]
\begin{center}
\caption{Parameters of the various models describing the DM distributions in various components of the MWG. The values of total mass, scale radius and central density for bulge, disk and halo are reproduced from Ref.~\cite{Sofue2013}. }
\vspace{3 mm}
\label{tab:mytable1}
\begin{tabular}{|c|c|c|c|c|c|c|c|c|}
\hline
\hline
Mass component & Total mass ($M_{\odot}$)      & Scale radius (pc)   & Central density  ($M_{\odot}$/pc$^3$)& Central pressure ($M_\odot$/{\rm pc {s}$^2$})\\
\hline
\hline SMBH           & $4.2\times 10^6$              & ---                   & --- &---\\

\hline Core   with Exp. Sph. & $4.2\times 10^{6}$              & $r_{co}=1.417\times 10^{-5}$ & $5.873\times 10^{19}$ &  $4.47$ \\

\hline Inner bulge    & $5.0\times 10^7$              & $r_{ib}$=3.8                   & $3.6\times 10^4$ & $1.21 \times10^{-19}$ \\

\hline Main bulge     & $8.4\times 10^9$              & $r_{mb}$=120                   & $1.9\times10^2$ & $3.36 \times10^{-21}$ \\

\hline Disk           & $4.4\times 10^{10}$           & $r_d=3\times10^3$         & $0.15$ & $3.35 \times10^{-24}$  \\

\hline Halo with NFW      & $5.0\times 10^{10} (r\leq h)$ & $h=12\times10^3$      & $1.1\times10^{-2}$ &    --- \\

 \hline
 \hline
  \end{tabular}
  \end{center}
\end{table*}

With the above information in our hands, we now focus on how to relate the theoretical RCs in GR and NG, to the measurements of the three-dimensional velocities of stars within our galaxy. We therefore proceed by taking into account different density profiles for each part of the MWG. In particular for the inner bulge, the main bulge, the extended disk and the halo we follow Ref. \cite{Sofue2013}. For the galactic core we either consider the presence of a SMBH or the case of a DM profile which can entirely replace the SMBH producing the same RCs for known stars.
Even though the presence and/or absence of a SMBH makes a big difference for a galaxy from other observational perspectives (such as jets, shadow, etc), our only intent here is to demonstrate that the observed rotation curves for stars near Sgr-A$^\star$ may be obtained from a DM profile without a SMBH and therefore at present we can not assert with certainty how the central mass of the galaxy must be divided between SMBH and DM.

\begin{table*}[ht]
\begin{center}
\caption{Parameters of the density profiles describing the halo of the MWG for the model with a DM distribution at the core without SMBH. For the Einasto profile the additional parameter as inferred from the best fit is given by $\alpha= 0.12$. We here adopt a simple analysis made by $\chi$-square and non-linear model fits developed through the software {\it Wolfram Mathematica}.} \vspace{3 mm}
\label{tab:mytable2}
\begin{tabular}{|c|c|c|c|c|c|c|c|c|c|c|c|}

\hline
\hline
Mass component & Total mass& Scale radius & Central density & Central pressure& BIC & AIC & $\Delta$BIC& $\Delta$AIC & $\chi^2$\\

 & ($ 10^{10} M_{\odot}$) &  ($10^{3}$ pc) & ($10^{-3}M_{\odot}$/pc$^3$) & ($10^{-27}M_{\odot}$/({\rm pc} {\rm s}$^2$))& & &  &  & \\

\hline\hline Halo with Isothermal  & $1.01 (r\leq h)$ & $h=12$      & $2.17$ & $9.01$ & $96.38$ & $95.99$ & 21.24  & 21.25 & 25.39 \\

  \hline Halo with Burkert& $3.06 (r\leq h) $& $h=12$ & $11.1$ & $1.81$ & $81.86$ & $81.46$ &6.72 & 6.72 & 5.06\\

    \hline Halo with Einasto & $17.5 (r\leq h)$ & $h=12$ &$20.1$ &$0.39$ & $81.65$ & $81.06$ &6.51& 6.32 & 4.26 \\

    \hline Halo with Exp. Sph. & $2.63 (r\leq h) $&$h=12$ & $ 7.56$& $53.11$ & $75.14$ & $74.74$ & 0 & 0 & 2.40\\

  \hline
  \hline
  \end{tabular}
  \end{center}
  \end{table*}

At first we consider the model involving a SMBH at the center of the galaxy and, following Ref.~\cite{Sofue2009,Sofue2012}, we reproduce known results.
The logarithmic plot of the measured rotation curves is shown in the left panel of Fig.~\ref{ris:image1ab} together with the observational data with error bars, covering the regions from the inner bulge to the halo.
The various curves represent the best-ﬁts with a central black hole of mass $M_{BH}=4.2 \times 10^{6} M_{\odot}$ \cite{Doeleman2008}, exponential sphere profiles for bulges and disk, and NFW profile for the halo. As mentioned earlier the NFW profile has a cusp at the center so it can not be used in models without the SMBH, therefore later we will consider other profiles for the halo.
The model parameters have been reproduced using the same ﬁtting procedure described in Ref.~\cite{Sofue2013,2020Galax...8...37S}.

Based on the above, the RCs for stars at a distance $r$ from the center is given by the sum of the individual contributions as
\be
v(r)^2=v_{BH}(r)^2+v_{ib}(r)^2+v_{mb}(r)^2+v_{d}(r)^2
+v_{h}(r)^2,\,
\label{eq:sample12}
\ee
where $v_{BH}(r)$, $v_{ib}(r)$, $v_{mb}(r)$ $v_{d}(r)$ and $v_{h}(r)$ are the linear velocities of test particles (stars) in the gravitational field of the BH, inner and main bulges, disk and halo, respectively. Using the least squares method, we reproduced and confirmed the results of Ref.~\cite{Sofue2013} including the error bars. The net theoretical rotation curve is shown in Fig.~\ref{ris:image1ab} (left panel) by a solid thick gray curve.

The BH mass is inferred by fitting the data within the range $10^{-3} \text{pc}\leq r\leq2 \text{pc}$.
At larger distances, the exponential sphere profile was used to fit the observed RC in the bulges of the MWG as
\be
\rho_{i}(r)=\rho_{0i}e^{-\xi_{i}(r)},\
\label{eq:denprofcore}
\ee
where $i=\{ib, mb\}$ and $\rho_i(r)$ is the volume mass density. The constant $\rho_{0i}$ is the central density, $\xi_{i}(r)=r/r_{0i}$ is the dimensionless radial coordinate and $r_{0i}$ is scale radius.
In general, for the above density profile $\rho_0$ and $r_0$ are free parameters that must be obtained from the observational data.

The expressions of the velocity for the disk and halo are described in detail in Ref.~\cite{Sofue2012}. Similarly, the parameters for the disk were calculated by fitting the data of the RC in the range $1\text{kpc}\leq r\leq 10 \text{kpc}$ and the parameters for the halo are found in the interval $10\text{kpc}\leq r\leq 400 \text{kpc}$, when considering the NFW profile.

In Table~\ref{tab:mytable1} we reproduce the main results of the fits for the whole MWG from Ref.~\cite{Sofue2013}. The curves in the left panel of Fig.~\ref{ris:image1ab} are constructed using the results in Table~\ref{tab:mytable1}.

As an alternative possibility we consider a model where the central black hole is replaced by an additional DM density profile $\rho_{co}(r)$. Correspondingly the contribution to the linear velocity due to the black hole $v_{BH}$ must be replaced by the corresponding one due to the core DM profile $v_{co}$.
As said, the NFW has a cusp at $r=0$ and so it can not be employed in models without a SMBH at the center. In addition it should be emphasized that due to large data errors (related to technical measurement difficulties) in the halo region discrepancies appear at large $r$ (see the left panel in Fig. ~\ref{ris:image1ab}). For this reason, we may consider different density profiles for the halo when modelling the galaxy without the SMBH. Specifically we fit the MWG data for the halo with the Burkert, Einasto, exponential sphere and isothermal sphere profiles. The various density profiles used for the halo are compared in the right panel of Fig.~\ref{ris:image1ab}.

Furthermore for the core, following Ref.~\cite{Boshkayev2019}, in the absence of a SMBH, the exponential sphere profile is used to model DM in the central parts of the galaxy.

\subsection{Bayesian analysis of halo density profiles}

What is the best profile for the halo? As said we do not consider the NFW profile because the presence of a cusp at $r=0$ is incompatible with the scenario where a DM core profile is present without a SMBH. Besides NFW and exponential sphere, the other commonly used profiles are the following~\cite{Cirelli2011}:
\begin{align}
&\text{Isothermal:} \hspace{2mm}    \rho_{Iso}(r)=\frac{\rho_0^{Iso}}{1+\xi_{Iso}^2}, \label{eq:isoprof}\\
&\text{Burkert:}   \hspace{6mm}      \rho_{Bur}(r)=\frac{\rho_0^{Bur}}{(1+\xi_{Bur})(1+\xi_{Bur}^2)},\\
&\text{Einasto:}   \hspace{6mm}      \rho_{Ein}(r)=\rho_0^{Ein} e^{2(1-\xi_{Ein}^{\alpha})/\alpha} ,
\label{eq:sample15}
\end{align}
where $\rho_0^{Iso}$ and $\rho_0^{Bur}$ are the central densities for the isothermal and Burkert profiles, respectively, $\rho_0^{Ein}$ is the characteristic density for the Einasto profile and we have defined  $\xi_{Iso}=r/r_0^{Iso}$,  $\xi_{Bur}=r/r_0^{Bur}$ and $\xi_{Ein}=r/r_0^{Ein}$, with  $\alpha$ the Einasto additional free parameter.

In Table~\ref{tab:mytable2} we present the parameters that best fit the various possible profiles for the halo of the MWG when the DM density profile at the core is used to replace the SMBH. For the inner bulge, main bulge and disk we use the same parameters shown in Table~\ref{tab:mytable1} in both models.

Notice that the exponential sphere is typically used for the inner parts of the galaxy, but we see here that one can apply it also to the halo, due to the large error bars in the halo and thanks to the fact that, unlike other profiles, it gives a finite mass for DM as $r\rightarrow \infty$.

In addition in Table~\ref{tab:mytable2}, we also show the statistical performances of the various profiles that may be used for the halo.
We considered two statistical criteria that are commonly used to quantify evidence\footnote{The statistical evidence can be in favor or against a given framework. In the first case, given a reference model, the confronted one works better than the reference. The second case is the opposite, {\it i.e.}, given a reference model, the confronted one is clearly disfavored.} of a given model against a reference one. The statistical criteria furnish two main information, {\it i.e.}, first they single out a reference model and second they quantify how the other models depart from the reference one.

The reference scenario is conventionally chosen as the one that minimizes the statistical criterion value among all the others. This is clearly a convention, one can select a different scenario as reference. In such a case, we could get negative differences, $\Delta$AIC and $\Delta$BIC. So, conventionally we decide to take such differences positive and to single out the reference paradigm as the one that minimizes the AIC and BIC values than all the rest. This choice does not mean that the reference paradigm is the absolute best background scenario but only \emph{the most suitable} compared with  the others. Moreover, these statistical criteria give us more information than the $\chi^2$ alone, since they also consider the number of free parameters that enter every single model under exam.

Hence, for our purposes we consider the widely-consolidated Akaike and Bayesian information criteria, more briefly AIC \citep{Akaike74} and BIC \citep{Schwarz78}, respectively. These two statistical rules read
\begin{align}
&\text{AIC}\equiv -2\ln \mathcal{L}_{max}+2p \ , \\
&\text{BIC}\equiv -2\ln \mathcal{L}_{max}+p\ln N\,,
\end{align}
where $\mathcal{L}_{max}$ is  the maximum likelihood estimate, $p$  the number of free parameters of each model and  $N$ the data points involved in our computation, as obtained from \citep{Sofue2013}.

Since the AIC and BIC hold their minimum values for the halo with exponential sphere mass profile, we selected this profile as reference model. All the other approaches suggest larger values for AIC and BIC and we thus report the corresponding numerical values in Table~\ref{tab:mytable2}. Moreover, we estimate the differences between those values and the minimum ones.

It is of utmost importance to quantify the differences of such quantities, namely $\Delta$AIC and $\Delta$BIC, since they are indicative of the goodness of each approach. To better understand this fact, let us summarize below the statistical meaning of such differences
\begin{itemize}\label{AICintervals}
  \item[{\bf 1.}] $\Delta\text{AIC(BIC)}\in [0,2]$ suggests only a weak evidence in support of the reference model.
  \item[{\bf 2.}] $\Delta\text{AIC(BIC)}\in (2,6]$ leads to a mild evidence in favor of  the reference model.
  \item[{\bf 3.}] $\Delta\text{AIC(BIC)}> 6$ implies a very strong evidence in favor of the reference model.
\end{itemize}
\noindent The reported values of Table~\ref{tab:mytable2} show that both $\Delta$AIC and $\Delta$BIC practically provide analogous outcomes. Also from Table~\ref{tab:mytable2} we see that there are huge discrepancies among possible models. In particular, both $\Delta$AIC and $\Delta$BIC show a strong evidence against the halo with isothermal profile. This means that the former mass component is mainly disfavored to fit the DM distribution in the halo.

The other mass components, namely halo with Burkert and Einasto profiles, are also disfavored, although less strongly than the isothermal profile. They could be therefore used for modelling the mass components, albeit they are evidently less statistically favored than halo with the exponential sphere.


\section{The dark matter equation of state}
\label{eos}

The different equations of state are obtained by solving the system of equations of hydrostatic equilibrium, for each portion of the galaxy with its own density profiles for DM. The free parameters of each profile are fixed from the fit of the observed data points of the galactic RCs.
For the core, we have considered the two options of a SMBH in vacuum and a core DM profile. As expected the difference in the influence on the dynamics at large distances for the two cases is negligible, so in the following we will consider only the exponential sphere profile profile for the core. For the inner bulge, main bulge and disk we also use the exponential sphere profile to describe the DM distribution. In particular, for the disc we considered the following equation \cite{Sofuepc}
 \be
     \rho_{d}(r)=\rho_{0d}e^{\left(\frac{1}{R_0}-\frac{1}{r_d}\right)r},
 \label{eq:sample14}
 \ee
where $r_d$ is the scale radius of the disk and $R_0=8$kpc is the distance of the Solar system from the galactic center. By redefining the new scale radius as $r_{0d}=R_0r_d/(R_0-r_d)$ the above equation can be written in the form of Eq.~\eqref{eq:denprofcore}. Therefore for all exponential sphere profiles we used Eq.~\eqref{eq:denprofcore} with different parameters $\rho_0$ and $r_0$.

\begin{figure*}[ht]
\begin{minipage}{0.49\linewidth}
\center{\includegraphics[width=0.97\linewidth]{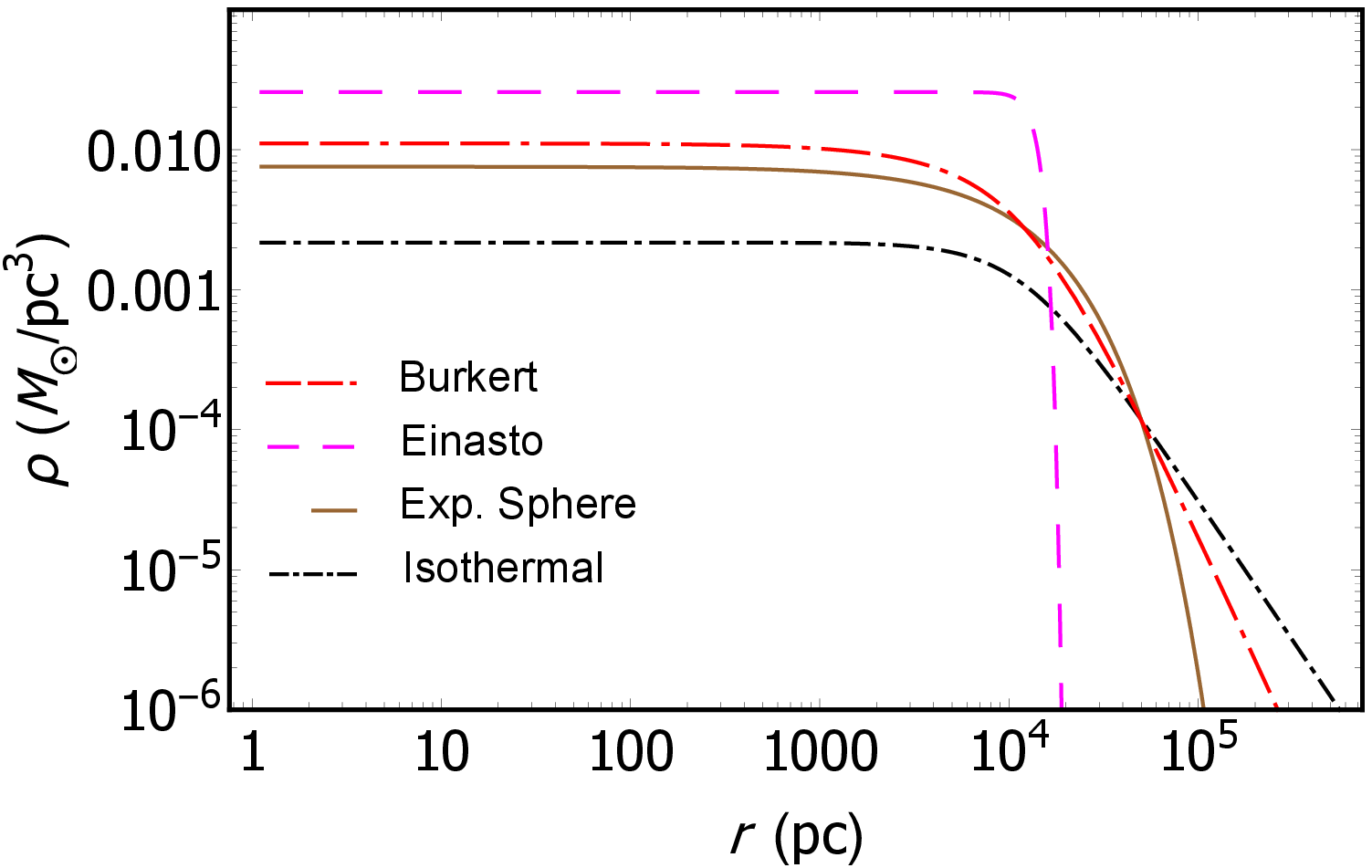}\label{ris:imagehalo0} \\ }
\end{minipage}
\hfill
\begin{minipage}{0.49\linewidth}
\center{\includegraphics[width=0.97\linewidth]{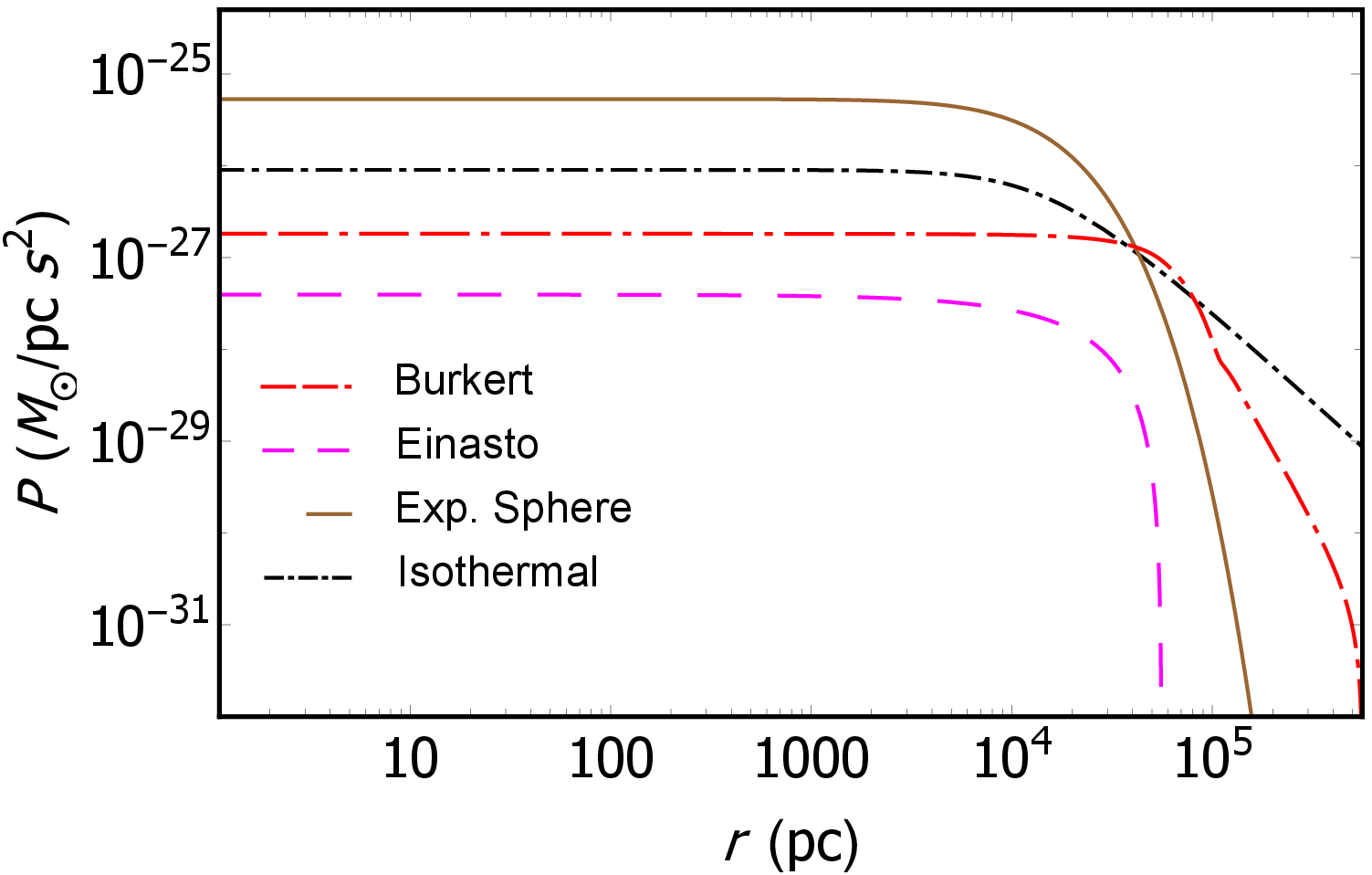} \label{ris:imagehalo1}\\ }
\end{minipage}
\caption{Color online. Left panel: Comparison of logarithmic density profiles of DM in the halo. Right panel: Comparison of the logarithmic pressure profiles of DM in the halo. For details see Table~\ref{tab:mytable2}}.
\label{ris:imagehalo2}
\end{figure*}
\begin{figure*}[ht]
\begin{minipage}{0.49\linewidth}
\center{\includegraphics[width=0.97\linewidth]{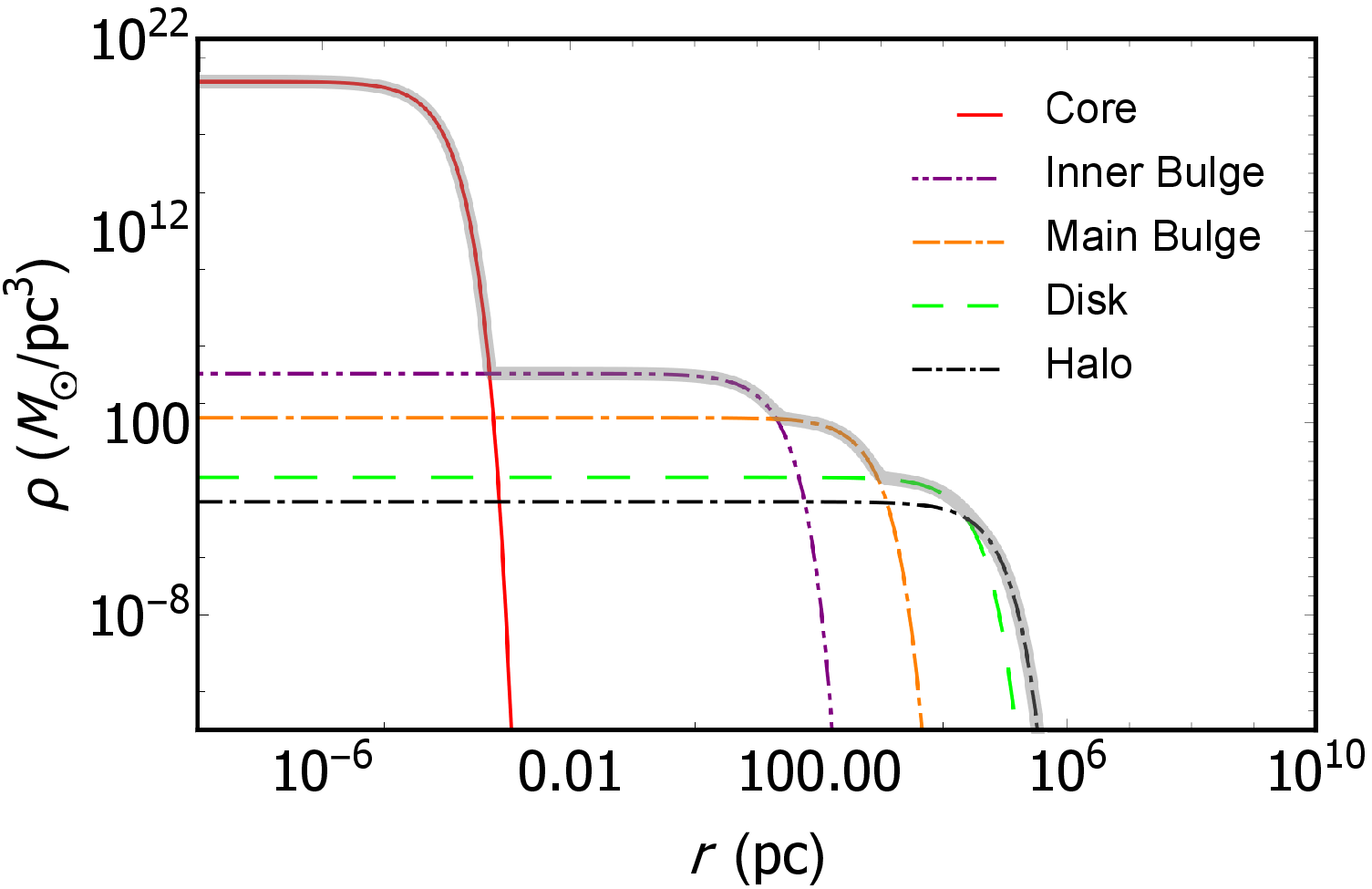}\label{ris:image4} \\ }
\end{minipage}
\hfill
\begin{minipage}{0.49\linewidth}
\center{\includegraphics[width=0.97\linewidth]{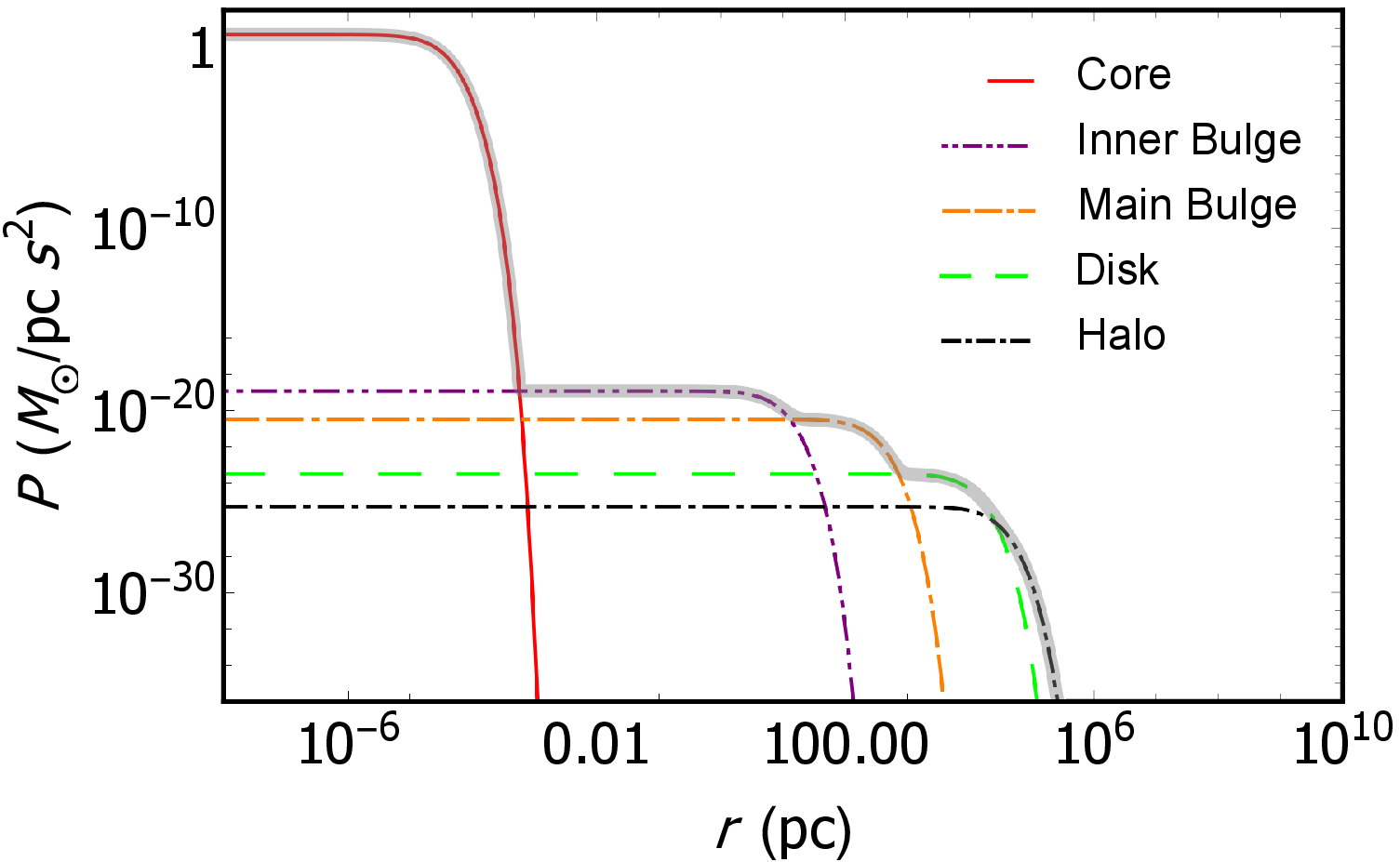} \label{ris:image5}\\ }
\end{minipage}
\caption{Color online. Left panel: Logarithmic density profiles of DM in the MWG. The DM distribution is divided into core, inner bulge, main bulge, disk, and halo, all modeled with the exponential sphere profile. Right panel: Logarithmic pressure profiles of DM in the MWG obtained from the density profiles. The net profile for both quantities in both panels is shown by a solid gray curve. }
\label{ris:image101}
\end{figure*}

In order to obtain the pressure profiles one needs to integrate Eqs.~\eqref{eq:sample6}-\eqref{eq:sample8} with the given density profiles and the appropriate boundary conditions
such that the pressure must vanish at the boundary of the cloud (which can be at spatial infinity). For the exponential sphere profile we get:
\bea \label{eq:sample17}
    P_i(r)&=& 8G\pi (r_{0i})^2 (\rho_{0i})^2 \left[
    \frac{1}{\xi_i}\left(e^{-\xi_i}-e^{-2\xi_{i}}\right)+\right. \nonumber\\
   &&\left.-\frac{1}{4}e^{-2\xi_i}- \Gamma(0,\xi_i)+ \Gamma(0,2\xi_i)\right],
\label{eq:sample18}
\eea
where the function $\Gamma$ is defined as $\Gamma(0,z)=\int^\infty_ze^{-t}t^{-1}dt$ and  $i=\{co, ib, mb, d, h\}$,  stands for the core, inner bulge, main bulge, disk and halo. Similarly we may solve the TOV equation for the other halo profiles considered above to obtain the corresponding pressures.
Note that the TOV equations are solved only numerically and at large distances from the center both NG and GR theories yield the same results.

The dependence of the density on the radial distance $r$ for the different profiles of the halo is shown in the left panel of Fig.~\ref{ris:imagehalo2} while the behavior of the corresponding pressure $P(r)$ is shown in the right panel. Notice that the Einasto profile exhibits a behavior considerably different from the rest at distances larger than $10^4$ pc.  Also notice that the differences between the profiles are not substantial, especially when related to the error bars in the available data. Therefore, keeping in mind the Bayesian analysis of the different profiles, in the following we employed the exponential sphere \eqref{eq:denprofcore} also for the description of the DM density profile in the halo.

In Fig.~\ref{ris:image101} the plots for the density $\rho(r)$ (left panel) and pressure $P(r)$ (right panel) of DM for the entire MWG are shown using the exponential spheres \eqref{eq:denprofcore},
with pressures given by \eqref{eq:sample18}. Of course, one can see that the pressure profile of the DM correlates with its density profile.

The speed of sound plays a key role in the perturbation theory to explain the formation of structures in the Universe \cite{Mukhanov1992}. Also the speed of sound determines the length above which gravitational instability overcomes radiation pressure and the perturbations grow. The existence of non-vanishing pressures for the DM fluid could also play a role at galactic scales. For an adiabatic fluid the speed of sound is defined as $c_s^2=\partial P/\partial \rho$.
In our model, the speed of sound for the core, inner bulge, main bulge, disk and halo is obtained from the exponential sphere profile and is given by \cite{galaxies2020}
\be
    \frac{(c_s^i)^2}{8\pi G}= r_{0i}^{2}\rho_{0i}\frac{(x_{i}-1)}{x_i\ln^2(x_i)}
     \left[1-\frac{(2+\ln(x_{i}))\ln(x_{i})}{2(x_{i}-1)}\right],
  \label{eq:sample143}
\ee
where $x_{i}=\rho_{0i}/\rho_i$.

\begin{figure}
\centering
\includegraphics[width=0.97\linewidth]{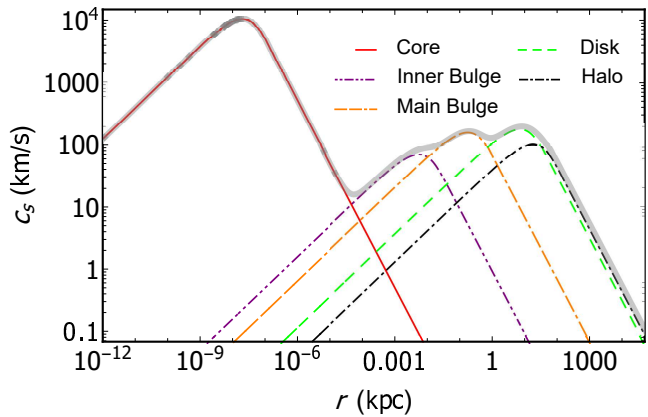}
\caption{The speed of sound $c_s$ as a function of the distance from the galactic center for the DM distribution in the MWG.}
\label{ris:imageT4}
\end{figure}

In Fig.~\ref{ris:imageT4} we show the speed of sound for the DM distribution in the MWG as a function of the radial distance from the center. It is interesting to note, that the behavior of the speed of sound is similar to the one of the rotation curve. Indeed, it can be seen from  Eqs.~\eqref{eq:sample3}, \eqref{eq:sample4} and \eqref{eq:sample9} that the relationship between $v(r)$ and $c_s(r)$ is given by
\be
c_s(r)=\sqrt{-\frac{d\ln r}{d\ln \rho(r)}\left(1+\frac{P(r)}{\rho(r)c^2}\right)}v(r).
\ee
Some sample values for the density, pressure and speed of sound of the galactic DM distribution at different distances from the center are given in Table~\ref{tab:newtable}. As expected, the speed of sound is larger towards the galactic center, where pressures are higher. Therefore if we wish to look for some observable effects of the DM EoS in the galaxy we should look closer to the galactic center.

\begin{table}[ht]
\caption{Density, pressure and speed of sound of the DM cloud for the MWG at various distances from the galactic center.}
\vspace{3 mm}
\label{tab:newtable}
\begin{tabular}{|c|c|c|c|c|}
\hline
\hline
Component & Distance  & Density  & Pressure & Sound speed \\
& (pc) & $\left(M_{\odot}/{\rm pc}^{3}\right)$ &$\left(M_{\odot}/{\rm pc} \hspace{0.5mm}{\rm s}^{2}\right)$ & $\left({\rm km/s}\right)$\\
\hline
\hline
Core & $10^{-6}$ & $5.5 \times 10^{19}$ & $4.4$ & 3771.7\\
\hline Inner bulge    & $0.1$ & $35065.0$ & $1.2\times 10^{-19}$ & 15.5\\
\hline Main bulge     &$150$ & $ 4.4$ & $ 1.3 \times 10^{-21}$  & $157.8$\\
\hline Disk           & $6 \times 10^{3}$ & $4.3 \times 10^{-2}$ & $1.3 \times 10^{-24}$ & $ 177.4 $ \\
\hline Halo & $3 \times 10^{4}$ & $6.2 \times 10^{-4}$ & $ 4.4 \times 10^{-27}$ & $92.7$\\
 \hline
 \hline
  \end{tabular}
\end{table}

Notice that the sound speed is not directly measurable. In principle, one can imagine to work out a back-scattering procedure, by measuring $v(r)$ and then if one knows the density either from observations or from theoretical considerations, then it is possible to get the sound speed. This would correspond to the velocity of perturbations of a DM system, i.e. quantifying the DM rate of propagation in galaxies, as a consequence of non-vanishing pressure. On the other hand it is also possible to postulate some classes of different DM EoS and then infer some bounds over the velocities and/or the sound speed at different radii. This would influence the whole analysis in galaxies and may be a direct signature of DM pressure, supporting the hypothesis that DM may not be under the form of dust.\footnote{This also would imply some consequences throughout the universe evolution, showing possible departures in the cosmological dynamics from early to late times.}

Further, a non-vanishing DM pressure clearly produces a \emph{dragging effect} on test particles that move inside the DM fluid. This is a fair consequence of DM background with a non-zero pressure as this effect happens for non-viscous fluids \cite{orlando0,orlando1,orlando3,orlando4,orlando5}. Using this prescription, a test particle average velocity within a DM background is proportional to the pressure gradient that appears in the presence of a non-zero pressure. This is valid in our scenario since we do not consider the effects of shear and/or heat exchange with the environment. Thus, differently from cosmological scales where any possible DM pressure gradient does not provide significant deviations in the overall potential and in the corresponding Hubble's flow, in such a case lying on galactic scales we expect significant departures especially close to the galactic center, while a decrease of effects outwards, {\it i.e.}, in far regions.

Since the effects of the DM profiles considered here are most relevant towards the center of the MWG we will now focus on the core and compare the two extreme scenarios, one with a SMBH in vacuum and the other with an exponential sphere profile for DM without the SMBH.
%
%
\section{Two models for the core}
\label{refindex}
We now focus our attention on the core of the galaxy and explore the differences between having a black hole in vacuum versus a dark matter cloud without the black hole. In order to do so we consider gravitational lensing due to the two gravitational fields. We see that noticeable differences between the two cases appear only at very short distances (smaller than 100 AU). Therefore at present, from the motion of S2 star, we can not exclude the possibility that Sgr-A* may be a black hole of smaller mass immersed in a dark matter cloud, in agreement with what discussed in Ref.~~\cite{2020A&A...641A..34B}.
Similarly, current measurements of light from the accretion disks at the center of distant galaxies may be due to the presence of both a supermassive black holes as well as a dark matter envelope and we can not put constraints to their mass ratio. One of the advantages of this scenario is that the total mass of the central region of a galaxy could be given by the sum of the SMBH mass and the DM mass, thus allowing for SMBHs in the early universe to be less massive than presently estimated while producing the same accretion disk spectra \cite{2020MNRAS.496.1115B}.


\subsection{Newtonian gravity versus general relativity}
\label{justification}
For most of the analysis in this paper we have considered Newtonian gravity (NG) since, as expected, the GR contributions are negligible at large distances. This is true also for the core of the galaxy. However, to give some quantitative statement on the limits of the Newtonian analysis, we can look at the differences appearing on the motion of test particles in the two cases.

More precisely, we consider the motion of test particles in the central part of the MWG (in the range from 0.25 to 50 AU) in the gravitational field of the BH without DM, and in the field of the DM core cloud without SMBH using the relativistic and classical formulas for the linear velocity (see Eqs.~\eqref{eq:sample10}-\eqref{eq:sample11b} and Table~\ref{tab:mytable3}).

In addition we can use the definition of the gravitational radius for the SMBH to formally define the corresponding radius for the DM distribution as
\be\label{eq:sample21}
r_{g}=\frac{2GM}{c^2},
\ee
where
\be\label{eq:sample34}
M=\begin{cases}
M_{BH}, \quad & \text{for BH},\\
M(r), \quad & \text{for DM},
\end{cases}
\ee
and $M(r)$ is the mass function of the DM distribution obtained from Eq.~\eqref{eq:sample6} using the density profile for the MWG core (see Eq.~\eqref{eq:denprofcore} and Table \ref{tab:mytable1}). Namely
\be
M(r)=\int^r_04\pi \tilde{r}^{2}\rho(\tilde{r})d\tilde{r}.\
\label{eq:massprof}
\ee
\begin{figure*}[ht]
\begin{minipage}{0.49\linewidth}
\center{\includegraphics[width=0.97\linewidth]{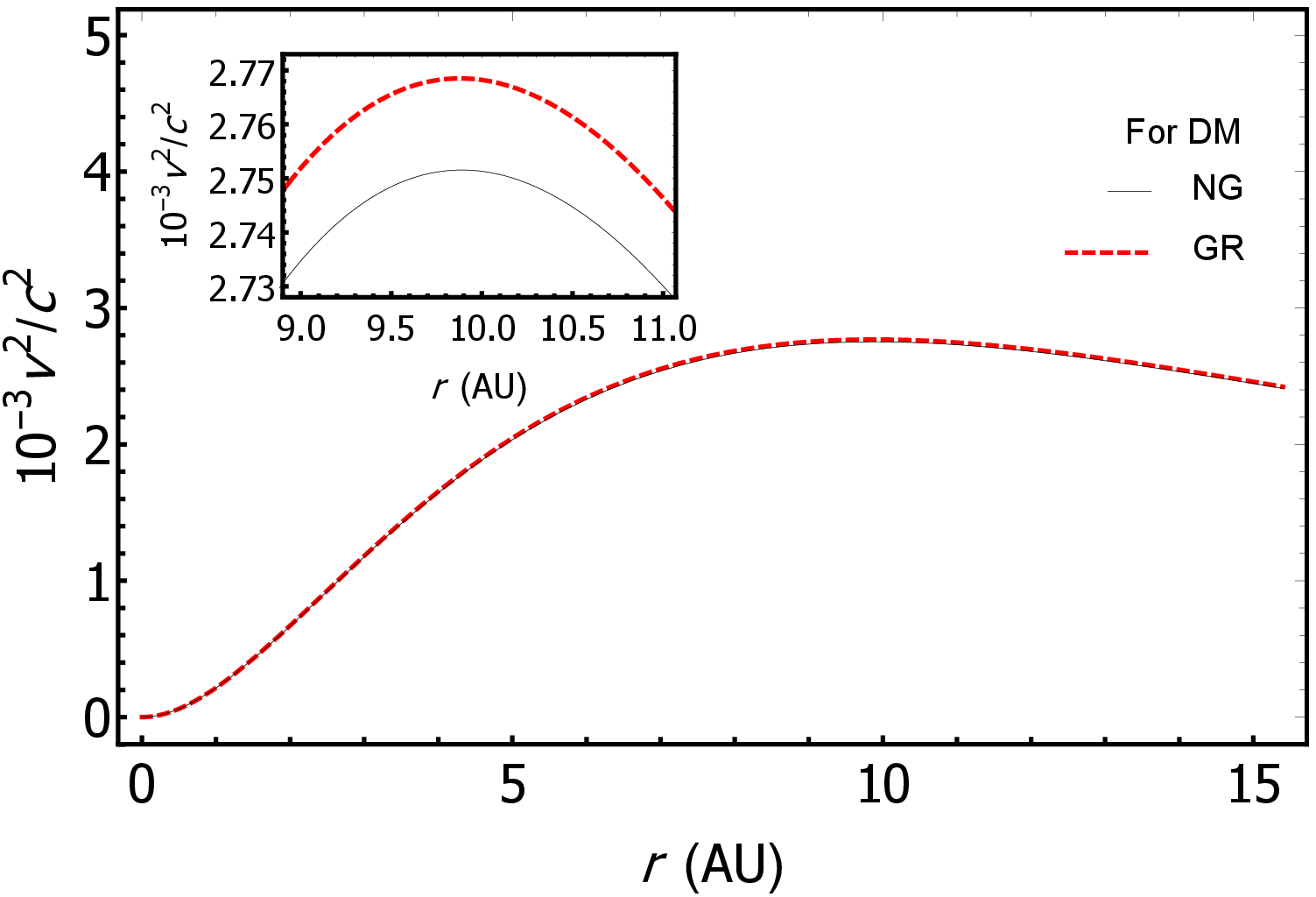}\\ }
\end{minipage}
\hfill
\begin{minipage}{0.50\linewidth}
\center{\includegraphics[width=0.97\linewidth]{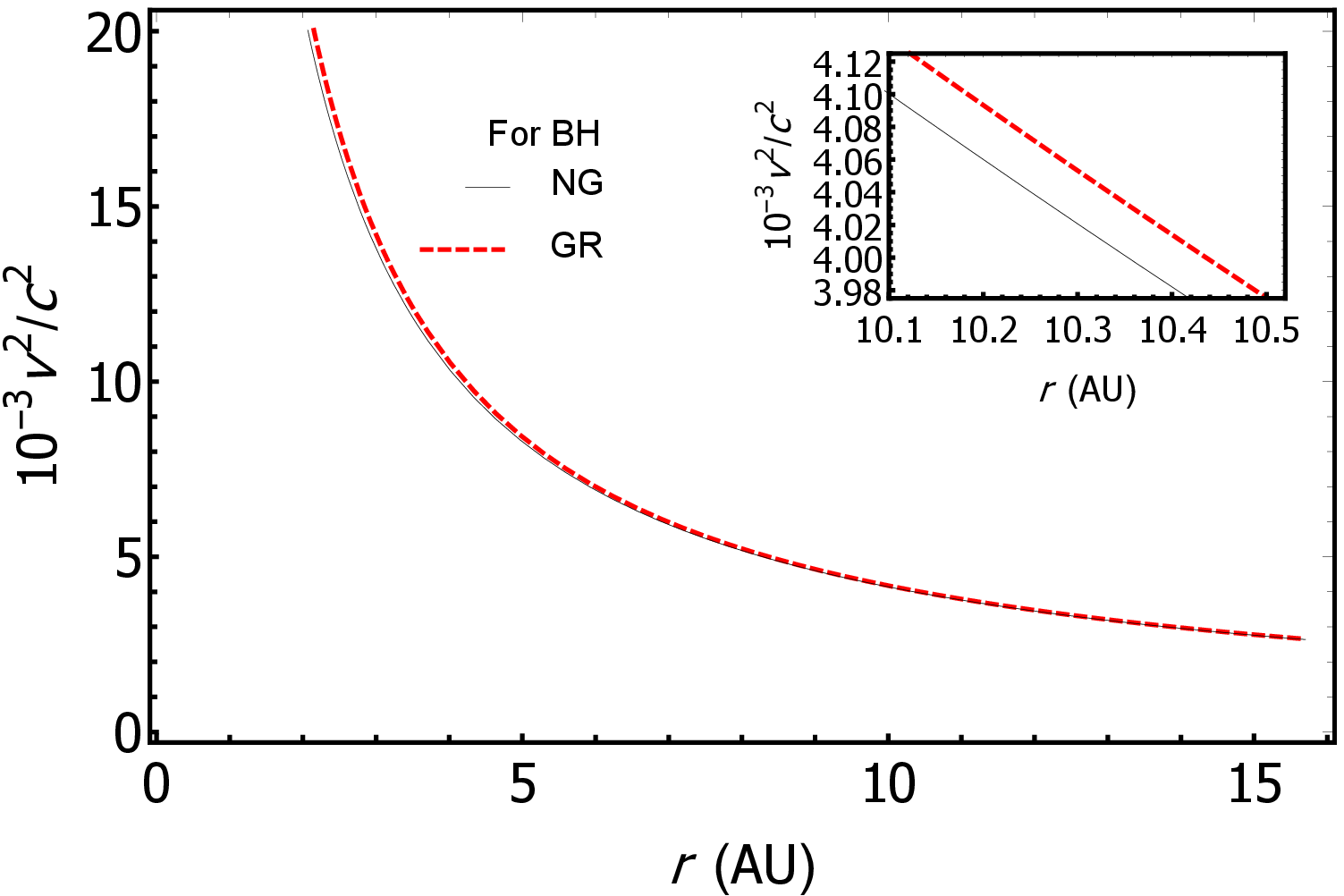}\\ }
\end{minipage}
\caption{Color online. Left panel: Dependence of the linear velocity of a test particle on the radial coordinate in the field of DM without SMBH, for circular orbits both in NG and GR. The solid black curve describes the function $v^2/c^2$ in NG and the dashed red curve in GR. Right panel: Dependence of the linear velocity of a test particle on the radial coordinate in the field of the SMBH without DM for circular orbits both in NG and GR. The solid black curve describes the function $v^2/c^2$ in NG and the dashed red curve in GR.
}
\label{ris:image43}
\end{figure*}
\begin{table*}[ht]
\begin{center}
\caption{Comparison of gravitational radius over radial distance $r_g/r$ for BH and DM, the speed of test particles $(v/c)^2$ in the fields of BH and DM, the gravitational potential $\Phi/c^2$ of BH and DM both in NG and GR at different distances $r$.}
\label{tab:mytable3}
\begin{tabular}{|c|c|c|c|c|c|c|c|c|c|c|}
\hline
   \hline $r$  & $10^{-3} r_g/r$ & $10^{-3} r_g/r$  & $10^{-3} \left(v/c\right)^2$   &  $10^{-3} \left(v/c\right)^2$  & $10^{-3} \left(v/c\right)^2$  & $10^{-3} \left(v/c\right)^2$ & $10^{-3} \Phi/c^{2}$ & $10^{-3}\Phi/c^{2}$ & $10^{-3}\Phi/c^{2}$ & $10^{-3}\Phi/c^{2}$\\
    (in AU) & for DM & for BH & DM in NG & DM in GR & BH in NG & BH  in GR & DM in NG & DM in GR & BH in NG & BH in GR\\
\hline
\hline 0.25 & 0.032  & 331.611& 0.01622  & 0.01626  & 165.805  & 248.066   & -7.083   & -7.115   & -165.805 & -201.442 \\
\hline 0.5  & 0.122  & 165.805& 0.06088  & 0.06105  & 82.902   & 99.380    & -7.060   & -7.091   & -82.903  & -90.644 \\
\hline 1    & 0.429  & 82.904 & 0.2146   & 0.2152   & 41.451   & 45.198    & -6.974   & -7.006   & -41.451  & -43.271 \\
\hline 2    & 1.337  & 41.451 & 0.669    & 0.671    & 20.725   & 21.621    & -6.694   & -6.724   & -20.726  & -21.167 \\
\hline 5    & 4.073  & 16.581 & 2.036    & 2.048    & 8.290    & 8.430     & -5.510   & -5.535   & -8.290   & -8.360 \\
\hline 10   & 5.508  & 8.290  & 2.754    & 2.771    & 4.145    & 4.180     & -3.778   & -3.791   & -4.145   & -4.162 \\
\hline 20   & 4.007  & 4.145  & 2.003    & 2.011    & 2.073    & 2.081     & -2.063   & -2.067   & -2.072   & -2.077 \\
\hline 30   & 2.757  & 2.763  & 1.379    & 1.382    & 1.382    & 1.385     & -1.381   & -1.383   & -1.382   & -1.384 \\
\hline 40   & 2.072  & 2.073  & 1.036    & 1.038    & 1.036    & 1.038     & -1.036   & -1.037   & -1.036   & -1.037 \\
\hline 50   & 1.658  & 1.658  & 0.829    & 0.830    & 0.829    & 0.830     & -0.829   & -0.830   & -0.829   & -0.830 \\
  \hline
  \hline
  \end{tabular}
  \end{center}
  \end{table*}

Then $M(r)$ is the amount of mass enclosed within a sphere of radius $r$ and is given by the integration Eq.~\eqref{eq:massprof} for a given density profile. Here we considered the exponential sphere in Eq.~\eqref{eq:denprofcore} and obtained
\be
M(r)=M_0F(\xi_{co}
(r)),\
\label{eq:sample37}
\ee
where $M_0=\lim\limits_{r\to \infty} M(r)=8\pi (r_{0co})^3\rho_{0co}$ is the total mass of the DM distribution in the region under consideration and
\be
F(\xi_{co})=1-e^{-\xi_{co}}\left(1+\xi_{co}+\frac{\xi_{co}^2}{2}\right),
\label{eq:sample38}
\ee
with $\xi_{co}(r)=r/r_{0co}$.

A simple numerical calculation shows that for the same total mass $M_0=M_{BH}$ given by the estimated mass of Sgr-A$^\star$ the difference between the mass function of DM $M(r)$ and the mass of the SMBH becomes less than 1\% at distances larger, than $11.65$AU. Consequently, 99\% of the mass of the DM core profile is concentrated inside a sphere of radius $11.65$AU at the center of the MWG.
In Table \ref{tab:mytable3} we compared the numerical values of the quantities $r_g/r$, $v^2/c^2$ and gravitational potential $\Phi$ for a test particle in the fields of the SMBH and DM.
\begin{figure*}[ht]
\begin{minipage}{0.49\linewidth}
\center{\includegraphics[width=0.97\linewidth]{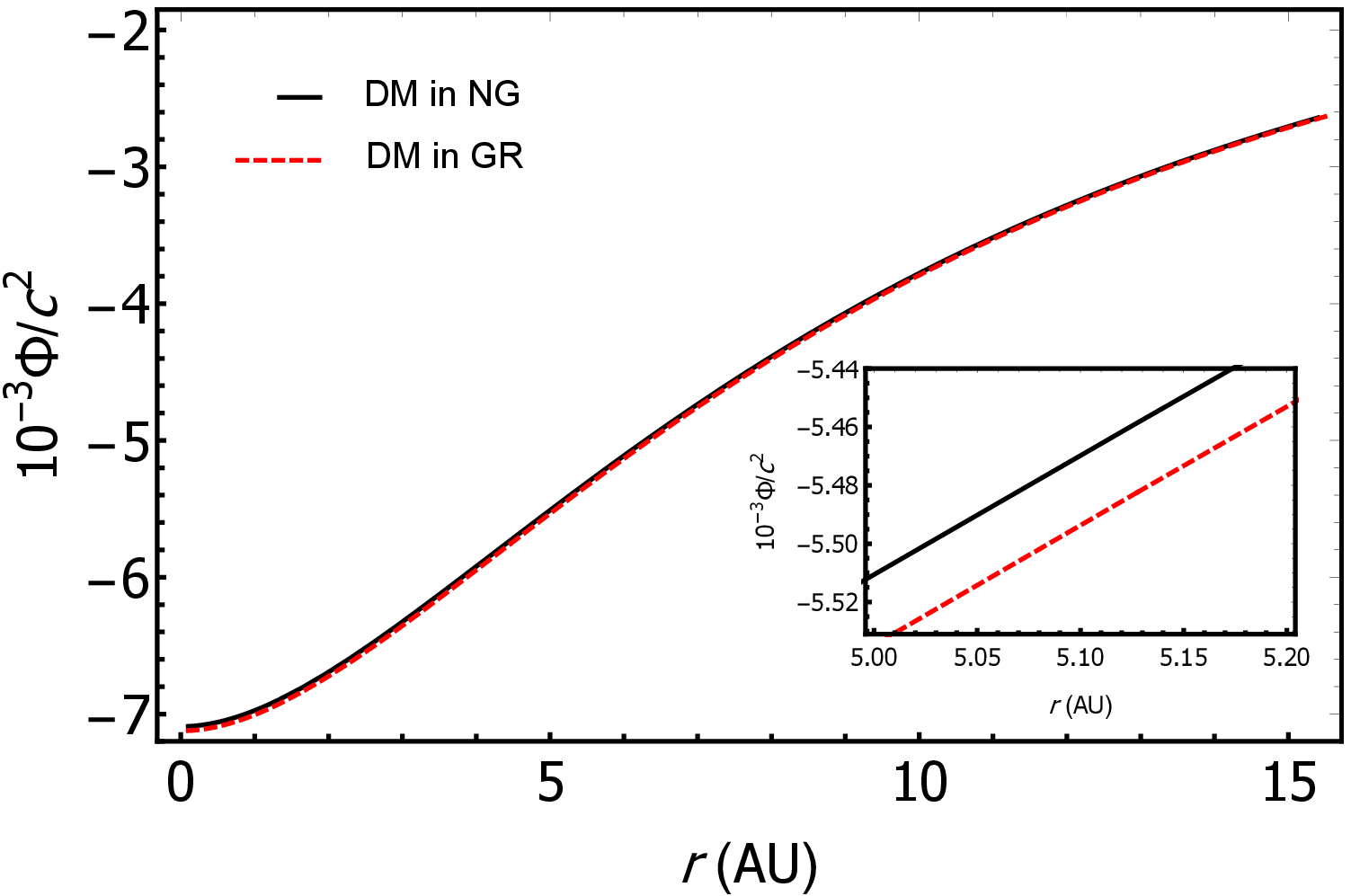} \\ }
\end{minipage}
\hfill
\begin{minipage}{0.50\linewidth}
\center{\includegraphics[width=0.97\linewidth]{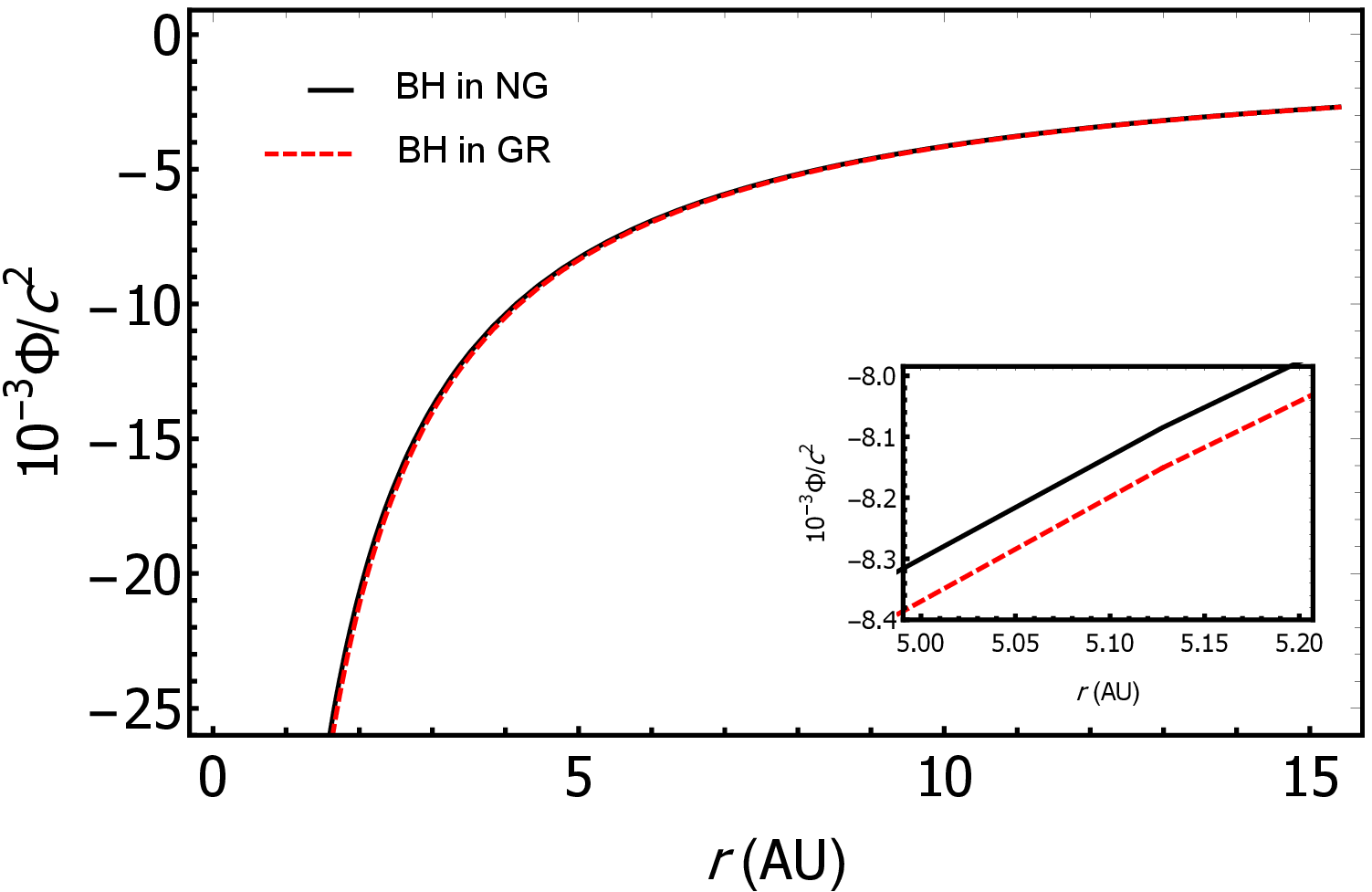} \\ }
\end{minipage}
\caption{Color online. Left panel: Comparison between the gravitational potential near the galactic center for the DM distribution in GR and NG. Right panel: Comparison between the gravitational potential of a BH in GR and that of a point particle of the same mass in NG.}
\label{ris:image2a}
\end{figure*}

\begin{figure}
\centering
\includegraphics[width=0.97\linewidth]{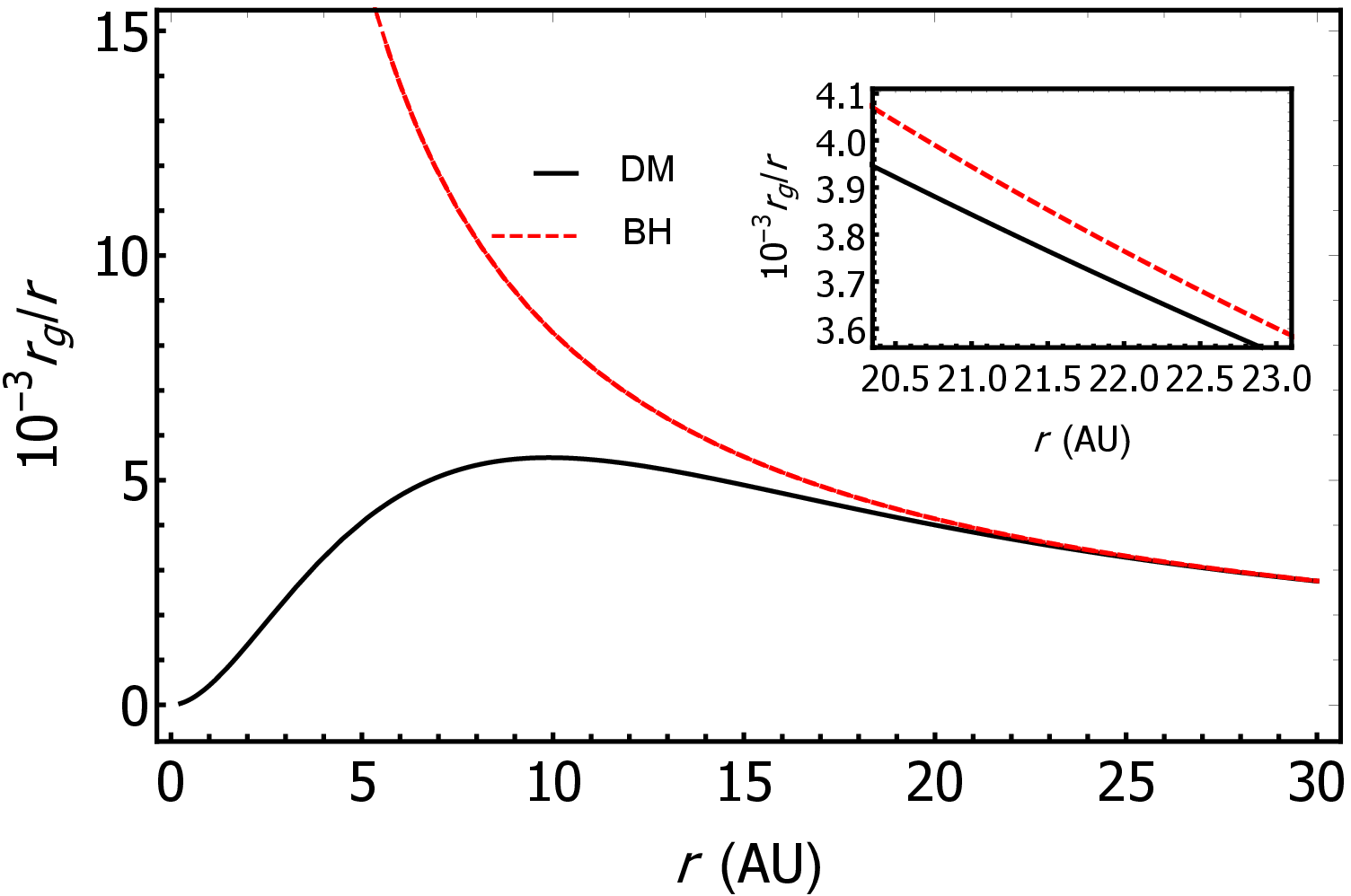}
\caption{Color online. Dependence of $r_g/r$ on $r$ for BH and DM. The dashed red curve corresponds to the case of a BH while the solid black curve corresponds to the DM profile. Notice that for $r<20$AU the two cases are significantly different.}
\label{fig:8}
\end{figure}

In the left panel of Fig.~\ref{ris:image43} we compared the velocity of a test particle in the gravitational field of the DM profile without SMBH in GR and NG using Eqs.~\eqref{eq:sample10} and \eqref{eq:sample10b}. It is easy to see that even at small distances (i.e. less than $10$AU) the differences are negligible.

Similarly, in the right panel of Fig.~\ref{ris:image43} we compared the velocity of a test particle in the field of the SMBH in GR and in the field of a point mass $M_{BH}$ in NG by making use of Eqs.~\eqref{eq:sample11} and \eqref{eq:sample11b}. Again it becomes clear that the differences between GR and NG appear only at short distances.

Similarly, in Fig.~\ref{ris:image2a} we compared the gravitational potential for DM (left panel) and BH (right panel) between GR and NG. It is clear that differences between the potentials in GR and NG appear very close to the galactic center at distances of the order of $10$AU and smaller (see Table~\ref{tab:mytable3} for details).

In addition, in Fig.~\ref{fig:8} we compared the radial dependence of $r_g/r$ from Eq.~\eqref{eq:sample21} in the case of DM with the case of the BH. It is clear that the differences between the two scenarios become important at about $20$AU, in a regime where NG is still a valid approximation.

We can summarize the comparison between the GR case and the NG case by stressing that for different values of the velocities $(v/c)^2$, gravitational radii over distance $r_{g}/r$ and gravitational potential $\Phi(r)/c^2$ we get different bounds on the departure of NG from GR:

\begin{itemize}

\item[-] For distances $r > 8.3$ AU, the relative error between the linear velocity $(v/c)^2$ of test particles in the field of the BH in GR and NG $[1-\left(v_{NG}/v_{GR}\right)^2]\times 100\%$ becomes less than 1 \%, so the Newtonian formula can be used from  $8.3$ AU and further.

\item[-] The relative error of velocities $[1-\left(v_{NG}/v_{GR}\right)^2]\times 100\%$ in the field of DM is always less than 0.63 \% at any $r$, this implies that one can safely use the Newtonian expressions for DM distribution, hence it is enough to study DM in NG.

\item[-] The relative error of BH gravitational potentials $[1-\Phi_{NG}/\Phi_{GR}]\times 100\%$ in NG and GR becomes less than 1\% at $r>4.16$AU.

\item[-] The relative error of DM gravitational potentials in NG and GR is always less than 0.45\% at any $r$.

\item[-] At $r\approx9.9$ AU, the maximum ratio of $r_{g}/r$ for DM is $5.5 \times 10^{-3}\ll1$, this means that the gravitational field of DM is relatively weak and so one can use NG (see. Fig. \ref{fig:8}).

\end{itemize}


\subsection{Lensing by dark matter}

We consider here the propagation of light in the gravitational fields of the DM distribution and compare it with that of a BH. In GR the equations of motion of a test particle are defined via the geodesic equations in a curved space-time, the curvature of which is generated by the energy-momentum tensor. The gravitational field of the galaxy is approximated by the static and spherically symmetric space-time metric \eqref{eq:sample1}. It is sometimes useful to convert the spherical coordinates used in the metric \eqref{eq:sample1} to isotropic coordinates \cite{Perlick2004} and rewrite the line element as follows
\be
\ ds^2=e^{2\Phi(\tilde{r})/c^2}\left[c^2dt^2-n(\tilde{r})^2({d\tilde r^2}+{\tilde r^2d\Omega^2})\right],\
\label{eq:sample22}
\ee
where the differential equation that relates isotropic coordinate $\tilde{r}$ to the spherical coordinate $r$ is
\be
\frac {d{\tilde{r}}}{dr}=\frac{\tilde{r}}{r\sqrt{{1}-\frac{2GM(r)}{c^2r}}},\
\label{eq:sample23}
\ee
and where we have introduced the scalar effective refractive index of a spherically symmetric gravitational field $n(\tilde{r})$ as
\be
\ n\left(\tilde{r}\right)=\frac{r(\tilde{r})}{\tilde{r}}{e^{\Phi\left(\tilde{r}\right)}}.\
\label{eq:sample24}
\ee
\begin{figure*}[ht]
\begin{minipage}{0.48\linewidth}
\center{\includegraphics[width=0.97\linewidth]{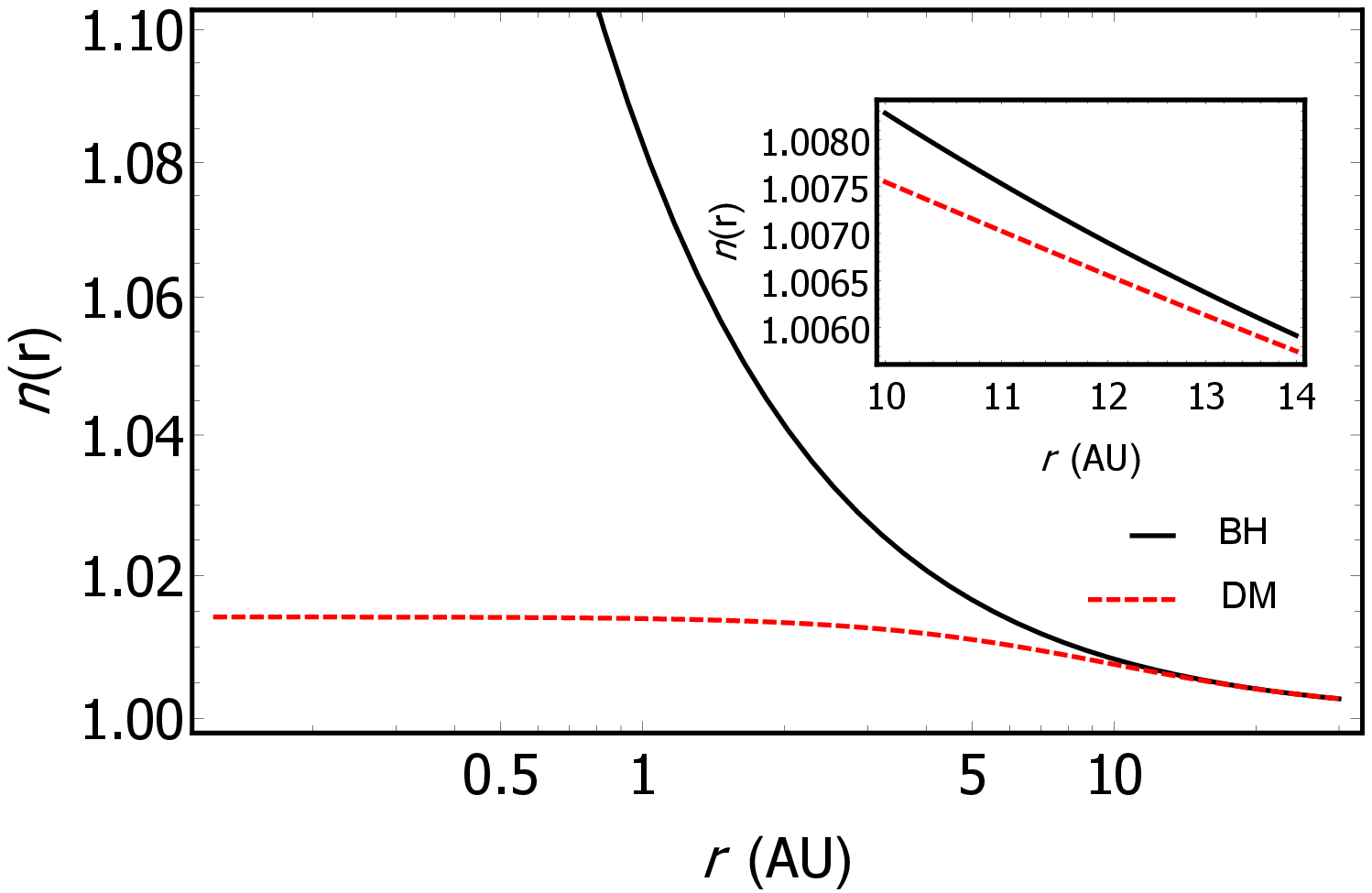}\label{ris:image7} \\ }
\end{minipage}
\hfill
\begin{minipage}{0.48\linewidth}
\center{\includegraphics[width=0.97\linewidth]{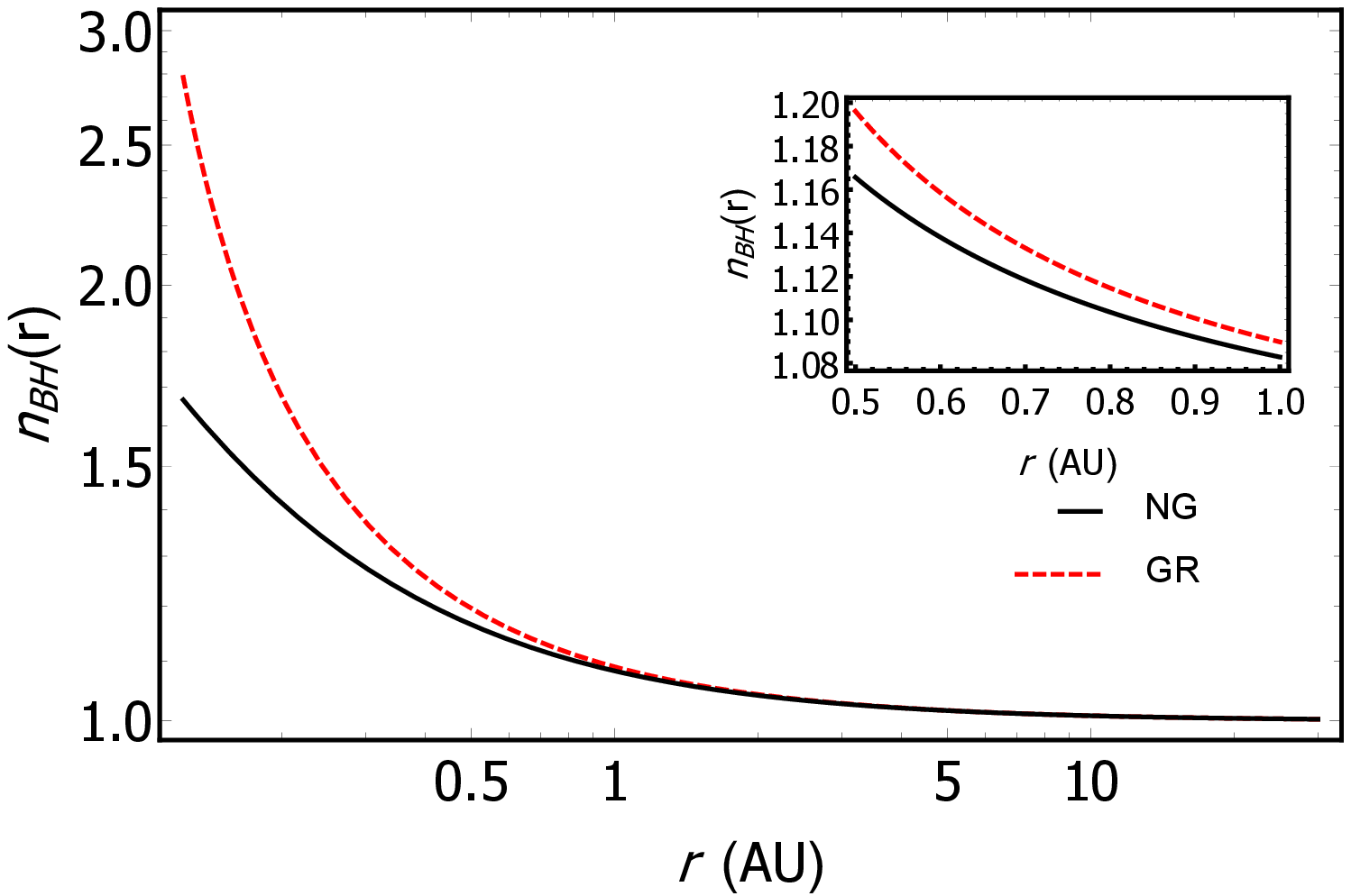} \label{ris:image8}\\ }
\end{minipage}
\caption{ Color online. Left panel: The refractive index of DM and BH. The DM is shown by a dashed red curve and BH is shown by a solid black curve in NG. Right panel: The refractive index in the field of BH in NG and GR. The refractive index is shown by a solid black curve in NG and by a dashed red curve in GR.}
\label{ris:image3ab}
\end{figure*}

Integrating equation (\ref{eq:sample23}), we find $\tilde{r}(r)$ as
\be
\tilde{r}=e^{\int\frac{dr}{{r}\sqrt{1-\frac{2GM(r)}{c^2r}}}},
\label{eq:sample25}
\ee
where the refractive index in the field of DM and BH in the weak field regime can be expressed as
\be\label{eq:sample26}
n({r})=\begin{cases}
\ {1}+\dfrac{2GM_{BH}}{c^2r}\, \qquad & \text{for BH},\\
\,\\
\ {1}-\dfrac{\Phi\left(r\right)}{c^2}-\displaystyle\frac{G}{c^2}\int_0^r\frac{ M\left(r^\prime\right)} {r^{\prime2}}dr^\prime\, \qquad & \text{for DM}.
\end{cases}
\ee
The refractive index is then calculated for the core of the MWG when there is a SMBH of mass $M_{BH}$ or DM distribution $M(r)$.

To switch variables from $r$ to $\tilde r$, we make use of Eq.
\eqref{eq:sample25}  and so, using Eq. \eqref{eq:sample26}, in Fig. \ref{ris:image3ab} (left panel) we plotted the dependence of $n(r)$ on $r$ in the gravitational fields of DM and BH. From here it is clearly seen that the refractive index of DM and BH can not be distinguished at distances starting from $r\simeq10$AU and larger.

Also, using Eq. \eqref{eq:sample24} we plotted the refractive index of the BH in NG and GR in Fig. \ref{ris:image3ab} (right panel). Here we see that the Newtonian description of the refractive index is good enough for distances larger than $r\simeq 1$AU.

Since the DM potential is smaller than the BH potential we can say that using the Newtonian description we may still be able to distinguish the two cases via lensing for observations in the range of approximately $1$AU to $10$AU. On the other hand, observations of gravitational lensing at distances larger than $10$AU from the center, could be described using NG but would not be able to tell the difference between the DM case and the BH case.

These results are in agreement with what was found by some of us in Ref.~\cite{2020MNRAS.496.1115B} where the refractive index in the gravitational field of a black hole surrounded by DM envelope was studied. The difference of the refractive index in the presence and absence of the DM envelope was small, but still not negligible.

Of course observations are not able yet to probe such distances from the central object either in the MWG or in other galaxies, but there is hope that future observations will allow to start putting some constraints on the possible existence of DM profiles at the center of galaxies and in particular improve our understanding of Sgr-A*.


\section{Final remarks}\label{concl}

We considered a model for the DM distribution in the MWG where the EoS for DM has non vanishing pressures and speed of sound. For the halo of the galaxy we considered different possible DM profiles and concluded that the exponential sphere provides a good fit for current observations. For the core of the galaxy we considered two extreme scenarios, one with a SMBH in vacuum and the other with an exponential sphere DM profile without the SMBH, with the idea that a more realistic situation would have the core's mass divided between DM and SMBH.

For other components of the galaxy (disk, main bulge, and inner bulge) the exponential sphere profile was also applied.

We solved the equations of hydrostatic equilibrium to obtain the relation between density and pressure of the DM profiles and consequently an estimate of the speed of sound in the DM fluid of the galaxy.

The impact that the non-vanishing EoS has on the motion of stars in the galaxy was also briefly discussed.

For the core of the galaxy, we showed that present observations either of the motion of S stars around Sgr-A$^\star$ or of the shadow of the central object in M87 are not sufficient to determine whether the object is a pure DM core or a SMBH in vacuum. Therefore, assuming that in a realistic scenario the core of a galaxy contains both a BH and DM, current observations do not allow to put constraints on the mass ratio of the two.
In particular, we evaluated the refractive index obtained on the basis of Fermat's principle, due to the SMBH or to the core's DM profile. We showed that in the case of the MWG the gravitational lensing effects due to a BH without DM or due to DM without a BH show appreciable differences only at distances of $10$AU or less from the galactic center.

This means that current observations of the movements of stars, such as S2, which exist in the vicinity of Sgr-A$^\star$, do not allow one to distinguish the proposed DM distribution from the SMBH. In view of recent observations of nearest stars to the galactic center \cite{2020ApJ...889...61P,2020ApJ...899...50P} it will be interesting to consider in future studies the motion of the star using our model in analogy to Ref.~\cite{2020A&A...641A..34B}.

\section*{Acknowledgements}
The work was supported by the Ministry of Education and Science of the Republic of Kazakhstan, Grant: IRN AP08052311. The authors express their gratitude to Prof. Y. Sofue for fruitful discussions on topics related to the disc profile and Dr. M. Muccino for technical assistance on statistical analysis employed in this paper.


\end{document}